\documentclass{pramana}

\usepackage{graphicx,amsmath,bm}
\usepackage{amsmath}
\usepackage{amssymb}
\usepackage{subfigure}
\usepackage{lipsum}
\usepackage[]{fancyhdr}
\usepackage{wrapfig}
\usepackage{float}
\usepackage{parskip}
\usepackage{lipsum}
\usepackage{mathtools}
\usepackage{cuted}
\usepackage{hyphenat}
\begin{document}

%%paper title
%%For line breaks \\ can be used within title 
\title{Observed Trends in FRB Population and Bi-modality in the Peak Luminosity Density Distribution}

%%author names are separated by comma (,) 
%%use \and before the last author name 
%%\textsuperscript{number} is used for affiliation
%%use a * along with the number separated by comma
%% for the  author for correspondence

\author{Nidhi Saini\textsuperscript{1,*}, Patrick Das Gupta\textsuperscript{1,2}}
\affilOne{\textsuperscript{1} Department of Physics and Astrophysics, University of Delhi, Delhi-110007}

%%escape two column mode for title, affiliation and abstract
%%by giving \twocolumn command as shown

\twocolumn[{

\maketitle

%%include \corres to print the corresponding author Email id
\corres{nsaini@physics.du.ac.in}

%%include \msinfo for
%%manuscript information such as
%%received, revised and accepted dates
%%
\msinfo{xxxxxxxxxxx}{ }{ }%{1 January 2015}{1 January 2015}{1 January 2015}

%%abstract
\begin{abstract}
Fast radio bursts (FRBs) are radio transients of extragalactic origin lasting for about a few to several milli-seconds. Their actual physical nature is still being actively researched on. In this paper, we have analyzed both non-CHIME and CHIME FRB data.  To circumvent the absence of measured fluence and flux density of FRBs belonging to the CHIME catalog, we have devised a novel approach that utilizes the ratio of the lower limits of the flux density $S_{\nu_O}$ to the fluence $F_{\nu_O}$ of individual FRB events to construct several parameters to investigate the presence of underlying trends in the FRB population drawn from both CHIME and non-CHIME data sets.  In this context, one of these parameters defined involve true brightness temperature as well as energy density, despite not knowing the actual size of the FRB emission region.

Our first robust conclusion is that the non-CHIME FRBs fall under two broad categories - those with luminosity density less than about $4\times 10^{33} $ erg/s/Hz at the frequency 300 MHz and those having larger luminosity density values than this. Our second robust result is that the parameters  computed using   $S_{\nu_O}/  F_{\nu_O} $ and other measured quantities are almost the same for both CHIME and non-CHIME FRB populations, vindicating our use of the technique based on the ratio of the lower limits of the flux density to the fluence.  This universality is also seen in the underlying patterns exhibited by the distributions of the computed parameters for both CHIME as well as the non-CHIME  FRB population, suggesting thereby the presence of the two luminosity density based categories  even  for the CHIME FRB population. Assuming that FRBs are caused by  magnetar glitches, we have discussed in this paper a simple physical model, incorporating an abrupt change in the light cylinder radius of an oblique rotator,  to address the existence of these two categories. 

%In the case of CHIME repeaters, distributions of a few of the dimensionless quantities hint at the existence of two modes of repeating radio transients.
\end{abstract}
%%insert keywords separated by comma using \keywords{words}
%%include \pacs{number} to print the PACS number
%{Radio Sources}
%\pacs{96.60.tg}%{Radio Emission}
}]
%%close the twocolumn escape here

%%include \doinum{number}for the DOI number in the header
%%include \volnum{number} for the volume number in the header
%%include \year{yyyy} for  year of publication in the header
%%include \pgrange{num--num} page range of article in the header
%%include \artcitid{num} for the article citation id
%%include \lp to print last page of the article
%%include \setcounter{page}{pagenum} for the exact starting page of the article

\doinum{xxxxxxxxxxx} %12.3456/s78910-011-012-3
\artcitid{\#\#\#\#}
\volnum{xxx}%{123}
\year{2024}%{2016}
\pgrange{xxxx}%{23--25}
\setcounter{page}{1}
\lp{19}%{25}

\footnotetext[2]{Superannuated on 31 March 2024}

\section{Introduction}
The first reported fast radio burst (FRB) was discovered serendipitously from the Parkes telescope archival data by  Lorimer \textit{et al}. (2007)~\cite{1}.  Since then numerous other FRBs have continued to be gleaned from different radio bands,  ranging from 110 MHz to about 8 GHz, by a host of radio telescopes. 

FRBs are luminous radio sources that are sporadic, and distant, appearing from random directions and lasting for about a few milliseconds. From the observed large dispersion measures (DMs) associated,  in general,  with these radio transients, their extra-galactic origin is almost certain.  While a majority of the FRB events are one-off affairs individually, several of the FRBs are repeaters including FRB 20121102A (the first ever repeater discovered~\cite{2}) located in a star-forming dwarf galaxy having a redshift $z=0.19$~(\cite{3}-\cite{5}).  The all-sky FRB event rate has been estimated to be fairly large $\sim 10^4$~\cite{6}.

Active radio transients from two of the repeaters - FRB 20121102A and FRB 20180916B, appear to be bunched with observed inter-bunch gaps of $\sim $ 157 days and $\sim$ 16 days, respectively, random occurrences within a bunch notwithstanding.  All the aforementioned points have been discussed comprehensively and thoroughly in many excellent reviews on the subject (e.g.~\cite{7}-~\cite{12}).  Because of the milli-second nature of the FRB durations, there are roughly two broad classes of models that are pursued seriously - one associated directly with magnetars (e.g.~\cite{6},~\cite{13}-\cite{20}) and the other, with the gravitational collapse of supra-massive spinning neutron stars~(\cite{21}-\cite{23}).

After the discovery of a low luminosity Galactic radio burst (FRB 20200428A)  along with an associated X-ray burst from a Soft Gamma Repeater (SGR) - the magnetar SGR 1935+2154~(\cite{24}-\cite{29}), the balance tilted somewhat in favor of the magnetar origin for at least a group of FRBs. However, the Five-hundred-meter Aperture Spherical Radio Telescope (FAST) has not seen any FRB-SGR connection so far, indicating that transient radio emission from magnetars during their soft gamma radiation phase is extremely rare~(\cite{30},\cite{10}). Furthermore, searches for high energy counterparts of FRBs have also led to low values of stringent upper limits on the flux densities of high energy photons~(\cite{31}-\cite{34}).

Despite a large body of research papers that continue to explore the theoretical modeling of FRBs, the real physical nature of these perplexing objects is still an enigma and is an area of active investigation. Therefore, it is very crucial to search for patterns that emerge from the observed FRB data to reach closer to the physical nature of FRBs. Some general trends that have ensued are that repeaters tend to have a larger observed duration than the non-repeaters and that the spectral width of the former is narrower leading to a wide variation in their spectral index values~(\cite{35},\cite{12}).

In this paper, by considering first the distribution of radio luminosity densities of non-CHIME FRBs, we present statistical evidence that suggests the existence of two categories of radio transients- a high luminosity density category and a lower one. The analysis assumed that most of the non-repeating FRBs have a spectral index value of 1.5 while the spectral index of various recurrences of radio transients from repeaters lie in a broad interval,  ranging from -2 to 1.5. 

Since only lower limits to the FRB fluence and flux density can be obtained from the CHIME catalog, we have introduced a new technique that makes use of dimensionless quantities that are computed using the ratio of the fluence lower limit to the flux density lower limit for each FRB and have carried out an investigation to demonstrate the existence of very similar statistical trends for both CHIME as well as non-CHIME FRBs. The observed universality in the values and distributions of these dimensionless quantities buttresses the validity of this technique.

\section{Basic Framework}

For a  transient radio source with a power law spectrum, the explicitly time-dependent intrinsic luminosity density may be expressed as,
\begin{equation}
\label{Lnu}%{eq.1}
 L_\nu (t)= L_0 (t)\ \nu^{-\alpha} \ ,    
\end{equation}
where $L_0(t)$,  $\alpha$, $\nu$, and  $t$ are the time-dependent source luminosity parameter, spectral index in the radio range, frequency, and time in the rest frame of the source,  respectively. 

In the case of a Friedmann-Lema$\hat{\mbox{i}}$tre-Robertson-Walker (FLRW) model, with the line element given by,
\begin{equation}
\label{FLRW}%{eq.1a}   
ds^2=c^2dt^2 - a^2 (t) \bigg[\frac{dr^2}{1-kr^2} + r^2(d\theta^2 + \sin^2{\theta } d\phi^2) \bigg ] \  ,
\end{equation}
the observed flux density corresponding to a cosmologically distant  extragalactic transient source then is given by,
\begin{align}
    S_{\nu_O} (t_O) &= \frac{(1+z) L_{\nu_O (1+z) }(t)}{4\pi D^2_L(z)} \label{Snu} \\%{eq.2} \\
    &= \frac{(1+z)^{1-\alpha} \nu^{-\alpha}_O L_0 (t)}{4\pi D^2_L(z)} \label{Snu1}%{eq.3}
\end{align}
    
where $z$,  $\nu_O$,  $t_O$ and $D_L(z)$ are the source cosmological redshift,  observed frequency, cosmic time in the observer's rest frame and the luminosity distance of the source, respectively, with time $t$  in the rest frame of the transient source being related to the observer's rest frame time $t_O$ by the equation,
\begin{equation}
\label{at0}%{eq.2a}
\frac{a(t_O)} {a(t)}= 1+z \ 
\end{equation}
and luminosity distance given by,
\begin{equation}
\label{Dlz}%{eq.4}
    D_L(z) = \frac{c}{H_0} (1+z) \int_0^{z} \frac{dz'}{\sqrt{\Omega_{\Lambda, 0}+\Omega_{m,0} (1+z')^3} } \ \ .
\end{equation}
In the subsequent analysis, to evaluate the luminosity distance from eq.(\ref{Dlz}), we have used the following values for the cosmological parameters pertaining to a flat $\Lambda$CDM model: $H_0=73.04$ km/s/Mpc, $\Omega_{m,0}=0.28$ and $\Omega_{\Lambda,0}=0.72$. 

The luminosity parameter $L_0(t)$ can be expressed in terms of the observed quantities using eq.(\ref{Dlz}),
\begin{equation}
 \label{L0t}%{eq.5a}  
 L_0(t)= \frac {4\pi D^2_L(z) \nu^{\alpha}_O S_{\nu_O} (t_O)} {(1+z)^{1-\alpha}}
\end{equation}

The observed fluence is given by,
\begin{equation}
\label{Fnu}%{eq.5}
    F_{\nu_O}= \int^{t_O + \Delta t_O}_{t_O} { S_{\nu_O} (t^\prime) \ dt^\prime} \ \ ,
\end{equation}
where $\Delta t_O = (1+z) \ \Delta t$ is the observed time duration of the source while $\Delta t$ is the intrinsic duration of the transient source in its rest frame. 

The energy released by the source is given by,
\begin{equation}
\label{Etot}
    E = \int^{\nu_2}_{\nu_1}{d\nu {\int^{t+ \Delta t}_t{L_\nu (\tau) d\tau}}} \ 
\end{equation}
$\nu_1$ and $ \nu_2 $ being the radio frequency range so that, after making use of eqs.~(\ref{Lnu}), (\ref{L0t}) and (\ref{Fnu}) in the above, we obtain, 
\begin{equation}
\label{Energy}%{eq.7}
    E=4\pi D^2_L(z) (1+z)^{\alpha -2} F_{\nu_O} \ \nu ^\alpha_O \bigg [\frac{\nu^{1-\alpha} _1 - \nu^{1-\alpha} _2} {\alpha-1} \bigg ]
\end{equation}

The redshifts of the repeating FRBs are estimated from the accurately measured redshifts of their host galaxies while the observed dispersion measures (DMs) are used to estimate the cosmological redshifts of the non-repeating FRBs.   The local contributions like from the source Doppler shift or gravitational redshifts to the overall redshift of a FRB are, in general,  negligible. 
For instance, if a source has a speed $\sim 300$ km/s  with respect to the cosmological rest frame, the corresponding Doppler shift is only $\sim 10^{-3}$. 

As far as  gravitational redshift is concerned,
emissions from around a massive and near-spherical  object of mass $M$ and size $R \geq 2GM/c^2$   would incur a  gravitational redshift $z_{gr}$ when observed at $r_{obs} \gg R$,
\begin{equation}
\label{zgr}%{eq.8}
z_{gr} (R) \cong \sqrt{\frac{1- \frac{2 G M} {c^2 r_{obs}}}{1- \frac{2 G M} {c^2 R}}} - 1 \approx \bigg (1- \frac{2 G M} {c^2 R} \bigg )^{-1/2} -1 \ \ .
\end{equation}
So, according to eq.~(\ref{zgr}), FRB radiation climbing out of the gravitational potential of a galaxy of mass $ \sim 10^{12}\ M_\odot $ and radial size $\sim 15$ kpc, would undergo a gravitational redshift $z_{gr} \sim 10^{-6}$, which is negligible. 

However, in the context of magnetar-centric FRB models,  if a radio transient is assumed to originate from a location close to the surface of a magnetar of mass $\sim 2\ M_\odot $ and radius $\sim 12$ km, the corresponding gravitational redshift is $\sim 0.4 $, which cannot be ignored. On the other hand, if the emission takes place at distances closer to the light cylinder $R_{lc}$ of the magnetar,
   \begin{equation}
   \label{Rlc}%{eq.9}
       R_{lc}= \frac{cP}{2\pi} \approx 50 \ \bigg (\frac {P} {10^{-3}\ s} \bigg ) \ \mbox{km} \ \ ,
   \end{equation}  
the gravitational redshift caused by even a milli-second magnetar's gravity is not very appreciable, since from eqs. (\ref{zgr}) and (\ref{Rlc}), $z_{gr}(R_{lc}) \cong 0.066 $ if the magnetar's mass and radius are $\sim 2\ M_\odot $ and $\sim 12$ km, respectively. Therefore, one may consider only the cosmological redshift $z$ in the analysis, as it is the dominant component.
 
In terms of the observed quantities, the source luminosity density at the rest frequency of 300 MHz can be obtained from eqs.(\ref{Lnu}) and (\ref{L0t}),
\begin{equation}
\label{L300}%{eq.10}
    L_{300} \equiv  4\pi D^2_L(z) (1+z)^{\alpha - 1} S_{\nu_O} \bigg (\frac{\nu_O}{300\ \mbox{MHz}} \bigg )^\alpha
\end{equation}
Analogously, the expression for the luminosity density at any specified frequency can be expressed.

\subsection{Brightness Temperature and Energy Density}

There is strong evidence that brightness temperatures of FRBs are very high, $(T_b)_{FRB} \gtrsim 10^{36}\ {}^\circ$K~(\cite{36},\cite{37}). In what ensues, we will consider the rest frame brightness temperature of a FRB (derived first by ~\cite{38}, in a somewhat different manner), for which the emission region has a radial size $\sim l_{em} $.  The brightness temperature  corresponding to the rest frame frequency $\nu $ is given by,
\begin{equation}
    \label{Tb}
    T_{b, FRB} = \frac{c^2} {2 k_B \nu^2} I_{\nu, FRB} \ \ , 
\end{equation}
where $ I_{\nu, FRB} $ is the rest frame specific intensity. 
\begin{figure}
    \centering
    \includegraphics[scale=0.25]{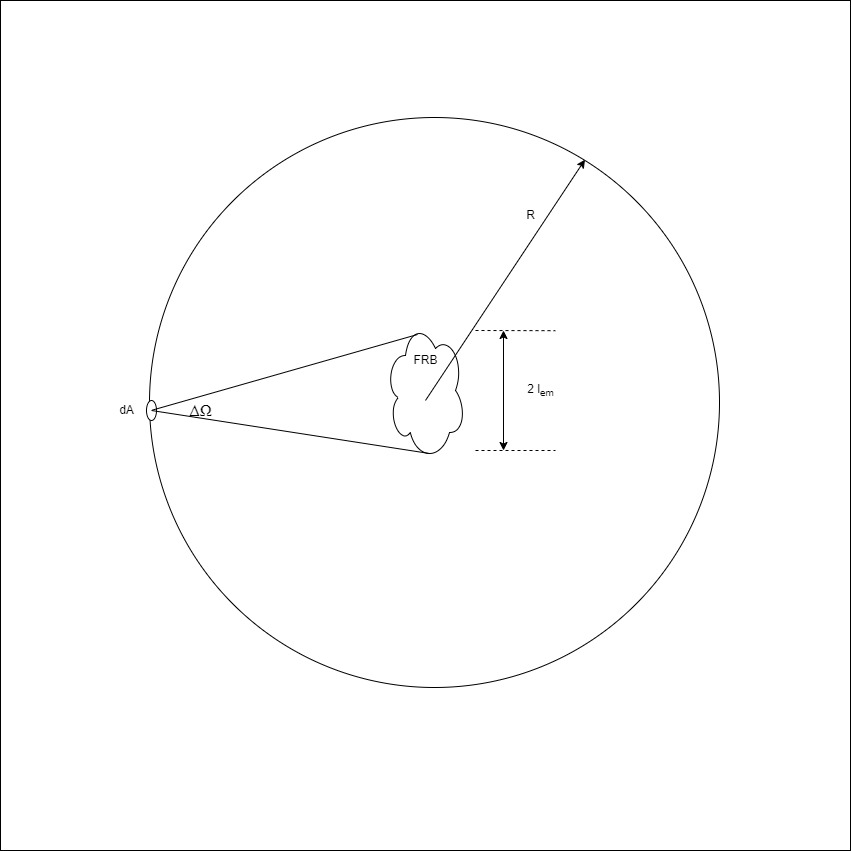}
    \caption{A schematic diagram for an FRB event}
    \label{fig:FRBevent}
\end{figure}
Since,
\begin{equation}
\label{Inu}
 I_\nu \ \Delta A \ \Delta t \ \Delta \Omega \ \Delta \nu = S_\nu \  \Delta A\  \Delta t\  \Delta \nu  \  ,
\end{equation}
at a radial distance $R \gg l_{em} $   from the FRB, that is within the galaxy hosting the FRB,   $ I_{\nu, FRB} $ is related to the rest frame flux density $S_\nu $ (see Fig (\ref{fig:FRBevent})),
\begin{equation}
\label{Inu1}
  I_{\nu, FRB} = \frac {S_\nu} {\Delta \Omega} = \frac{R^2 \ S_\nu}{\pi l^2_{em}} \ \ . 
\end{equation}
As the observation point is at a cosmological distance from the FRB, to express  $T_{b, FRB}$ in terms of the measured flux density $S_{\nu_O}$ at the observed frequency $\nu_O$, we may use the total number of photons $\Delta N$ emitted by the FRB in the frequency interval $(\nu, \nu + \Delta \nu)$ and  the time interval $(t, t+\Delta t)$,
\begin{align}
    \label{DelN}
    \Delta N &= 4 \pi R^2 \ S_\nu \ \Delta \nu \ \Delta t / h \nu \\
        &= 4 \pi a^2_0\ r^2 S_{\nu_O}\  \Delta \nu_O  \ \Delta t_O / h \nu_O \label{DelN1}\ \ ,
\end{align}
where $a_0$ and $r$ are the present-day expansion scale factor and the Robertson-Walker radial coordinate of the FRB, respectively.

Using the well known relations,  luminosity distance $D_L(z) = a_0 (1+z) r $, $\Delta t_O =(1+z)\ \Delta t$ and $\Delta \nu_O =\Delta \nu/(1+z)$ in eqs.(\ref{DelN}) and (\ref{DelN1}), we obtain the required relation,
\begin{equation}
    \label{RSrel}
    R^2 \ S_\nu = D^2_L(z) \bigg ( \frac  {S_{\nu_O}}  { (1+z) } \bigg )\ \ ,
\end{equation}
so that after making use of eqs.(\ref{Inu1}) and (\ref{RSrel}) in eq.(\ref{Tb}),  the rest frame brightness temperature   is of the form,
\begin{align}
  \label{TB}
  T_{b, FRB} &= \frac{c^2} {2 \pi k_B \nu^2_O}  \bigg (\frac{S_{\nu_O} } {l^2_{em}} \bigg ) \bigg (\frac{D^2_L (z) } {(1+z)^3} \bigg ) \\ &= \frac{c^2} {2 \pi k_B \nu^2_O} S_{\nu_O} (1+z) \bigg (\frac{  D^2_A (z)} {l^2_{em}} \bigg )   \ ,
\end{align}
where $D_A(z)= D_L(z)/(1+z)^2$ is the angular diameter distance.
%where $D_A = D_L/ (1+z)^2$  is the angular diameter distance.    

From  special relativity and causality arguments, it follows that  $l_{em} \lesssim  c \Delta t $ so that the brightness temperature of the source in its rest frame  (eq.(\ref{TB})) satisfies an inequality,
%Pawan Kumar and ...2018;   Lou et al. 2023; Zhang review 2023],
%\begin{equation}
%\label{eq.11}
 %   (T_b)_{FRB} \cong \bigg ( \frac{c^2}{2 \pi k_B \nu^2_O} \bigg ) S_{\nu_O} (1+z) \bigg ( \frac{D_A (z)}{l_{em}} \bigg )^2 
%\end{equation}
\begin{align}
\label{TbFRB}
   T_{b,FRB} &\gtrsim T_b \equiv \bigg ( \frac{c^2}{2 \pi k_B \nu^2_O} \bigg ) S_{\nu_O} (1+z) \bigg ( \frac{D_A (z)}{c \Delta t} \bigg )^2 \\
     &= \frac{ S_{\nu_O} D^2_L(z) (1+z)}{2 \pi k_B \bigg(\Delta t_O \ \nu_O (1+z)\bigg)^2} \label{TbFRB1} \ \ ,
\end{align}

The lower limit $ T_b $ to the actual brightness temperature given by eq.(21) has been thoroughly discussed by Luo \textit{et. al.} and Zhang  ~(\cite{38},\cite{12}).

%and the observed duration $\Delta t_O= \Delta t (1+z) $.

The relativistic causality considerations can also be used to estimate a lower bound to the energy density of the radio photons within the emitting region,
\begin{equation}
\label{uFRB}
  u_{FRB}  \cong \frac{E}{l^3_{em}} \gtrsim u \cong \frac{E}{(c \Delta t)^3} 
\end{equation}
The energy density $u_{FRB}$ is likely to be associated with the magnetic field as well as the spin angular momentum of the compact object.

\subsection{FRB Parameters constructed using measured quantities and ratio of flux density to fluence}
It is useful to define dimensionless as well as dimensional quantities that make use of various measured and estimated physical quantities related to the FRBs so that we may compare the obtained outcomes for both non-CHIME as well as CHIME FRBs. Since the CHIME catalog only provides lower bounds to the fluence and flux density of individual radio transients, we have introduced a new technique that involves taking the ratio of fluence to flux density and vice versa. 

The rationale for adopting  this strategy  is that  the flux density and the fluence of an individual FRB event  recorded by CHIME over milli-second scale intervals, would both be underestimated by approximately the same factor implying thereby,

$$ \frac {(S_{\nu_O})_\text{lower bound}}  {(F_{\nu_O})_\text{lower bound}} \cong \frac{{(S_{\nu_O})_\text{measured}} }{ {(F_{\nu_O})_\text{measured}}  } \ \ ,$$
since the direction of an FRB as seen by the CHIME telescopes does not change appreciably in such a short interval.

As the FRB energy $E$ (eq.(\ref{Energy})) and the luminosity density $L_{300}$ (eq.(\ref{L300})) are proportional to fluence and flux density, respectively, we have defined various parameters below that can be computed for CHIME FRBs too in a meaningful manner (except for the quantity $X5$, which is a dimensionless characterization of the energy density $u$ and can be computed only for non-CHIME data).

\begin{align}
\label{X1}
    X1 &\equiv  \frac{E \ \ L_{300}}{(F^2_{\nu_O} +S^2_{\nu_O} \Delta t^2_O) \nu_O^2 (\frac{DM}{n_e})^4} \\
    &= \frac{E \ \ L_{300}}{F^2_{\nu_O}(1 + \frac{S^2_{\nu_O}} {F^2_{\nu_O}} \Delta t^2_O) \nu_O^2 (\frac{DM}{n_e})^4}
\end{align}

\begin{align}
\label{X2}
    X2 &\equiv \frac{E \ \ L_{300}}{(F^2_{\nu_O} +S^2_{\nu_O} \Delta t^2_O) (\frac{DM}{n_e})^2 c^2} \\
    &= \frac{E \ \ L_{300}}{F^2_{\nu_O}(1 + \frac{S^2_{\nu_O}} {F^2_{\nu_O}} \Delta t^2_O)  (\frac{DM}{n_e})^2 c^2}
\end{align}

\begin{align}
\label{X3}
    X3 &\equiv \frac{L_{300}}{\sqrt{(F^2_{\nu_O} +S^2_{\nu_O} \Delta t^2_O)} \ \nu_O (\frac{DM}{n_e})^2} \\
    &= \frac{L_{300}}{ F_{\nu_O} \sqrt{(1 + \frac{S^2_{\nu_O}} {F^2_{\nu_O}}  \Delta t^2_O)} \ \nu_O (\frac{DM}{n_e})^2}
\end{align}

\begin{equation}
    X4 \equiv \frac{u}{L_{300}} . \big(\frac{DM}{n_e}\big)^3 \label{X4} %; \  X5 \equiv \frac{u \ \ c^3}{h \nu^4_O}\ ; \   X6 \equiv \frac{k_B\ T_b}{E} 
\end{equation}

\begin{equation}
    X5 \equiv \frac{u \ \ c^3}{h \nu^4_O} \label{X5}
\end{equation}

\begin{equation}
    X6 \equiv \frac{k_B\ T_b}{E}  \label{X6}
\end{equation}

\begin{align}
    X7 &\equiv \frac{k_B T_{b,FRB}} {u_{FRB}^{2/3} E^{1/3}} \label{X7} \\
     &= \frac{c^2}{2 \pi \nu_0^2} \bigg(\frac{S_{\nu_0}}{E}\bigg) \bigg( \frac{D^2_L(z)}{(1+z)^3}\bigg)
    %\frac{c^2}{8 \pi ^2} \bigg(\frac{1}{\nu^{\alpha +2}}\bigg) \bigg(\frac{S_{\nu_0}}{F_{\nu_0}}\bigg) \bigg(\frac{\alpha -1}{\nu_1^{1-\alpha}-\nu_2^{1-\alpha}}\bigg) \bigg(\frac{1}{(1+z)^{\alpha +1}}\bigg) \label{X7}
\end{align}

In the expressions for $X1$ to $X4$, $n_e$ that appears is a fiducial number density of electrons used to obtain a `length' dimension from the DM. In this paper, we have set $n_e= 0.03\ \mbox{cm}^{-3}$, which is the mean value of the Galactic disc electron number density. 

$X1$ and $X2$ are  distinct dimensionless parameters  characterizing the product of the FRB energy and luminosity density in two different ways, while $X3$ is a dimensionless parameter associated with the luminosity density alone. The quantity $X4$, on the other hand, is a dimensionless parameter capturing the essence of the the physical quantity $u/L_{300}$. A lower bound to the degree of coherence per radio photon is characterized by the dimensionless parameter  $X6$.

The parameter $X7$ defined by eq.(34), involving the actual brightness temperature given by eq.(21) and the energy density $E/l^3_{em}$, is so constructed that only the measured and estimated quantities as well as the ratio of flux density to fluence come into play (eq.(35)) in computing the parameter, despite not knowing the FRB size $l_{em}$.  Therefore, $X7$ can be estimated for the CHIME FRBs.

%\begin{equation}
%    X5 \equiv \frac{u \ \ c^3}{h \nu^4_O}
%\end{equation}
%\begin{equation}
%    X6 \equiv \frac{k_B\ T_b}{E}
%\end{equation}

\section{Data} \label{data} The non-CHIME FRB data analyzed in the present work is obtained from the Transient Name Server (TNS) website $https://www.wis-tns.org/$.  A majority of these transient events were detected using telescopes like the Australian Square Kilometre Array Pathfinder (ASKAP), Parkes, Arecibo, etc.  While the  Canadian Hydrogen Intensity Mapping Experiment (CHIME)  data is taken from the CHIME/FRB first catalog paper from the CHIME data website $https://www.chime-frb.ca/catalog$~\cite{39}.

The redshift information for non-CHIME one-off FRBs has been taken from FRBSTATS~\cite{40} and the corresponding data for non-repeaters (NRs) in the CHIME catalog from Tang \textit{et. al.}~\cite{41}. We have also used the individual sub-burst data of FRB 20121102 ~(\cite{2}-\cite{5},~\cite{42}), and each of these sub-bursts has been treated as distinct bursts in our study, and have been classified as repeaters (REPs). We have adopted this classification for every repeater (e.g. sub-bursts of FRB 180916.J0158+65~\cite{43}).  

\section{Analysis and Results} Spectral index for NRs is taken to be 1.5 throughout (this is found to be true for 23 ASKAP FRBs~\cite{44}), except in the cases of FRB 20070724A and FRB 20110523A for which the spectral indices are taken to be 4 and 7.8, respectively~(\cite{1},\cite{45}). The spectral indices for REPs vary over a wide range, presumably because of their narrow spectral width [35,12]. For the calculations of energy as well as luminosity density (eqs.~(\ref{Energy}) and ~(\ref{L300})) and subsequent analysis, we have considered five independent trials in each of which a set of values of $\alpha \in  $ $ \lbrace -2, -1.5, -1.3,-0.5, 1.3, 1.5 \rbrace $  are randomly chosen, using a uniform probability distribution, and assigned to the REPs. Thereafter, we have also considered another trial and in this sixth trial, all the REPs are assigned $\alpha = 1.5$. The same procedure has been carried out for each of the recurrences of a CHIME repeater too. (For the constraints on the values of $\alpha$ see~(\cite{46}-\cite{49})). 

\subsection{Non-CHIME} 
\begin{figure}[!ht]
    \centering
    \subfigure[]{\includegraphics[width=0.44\textwidth]{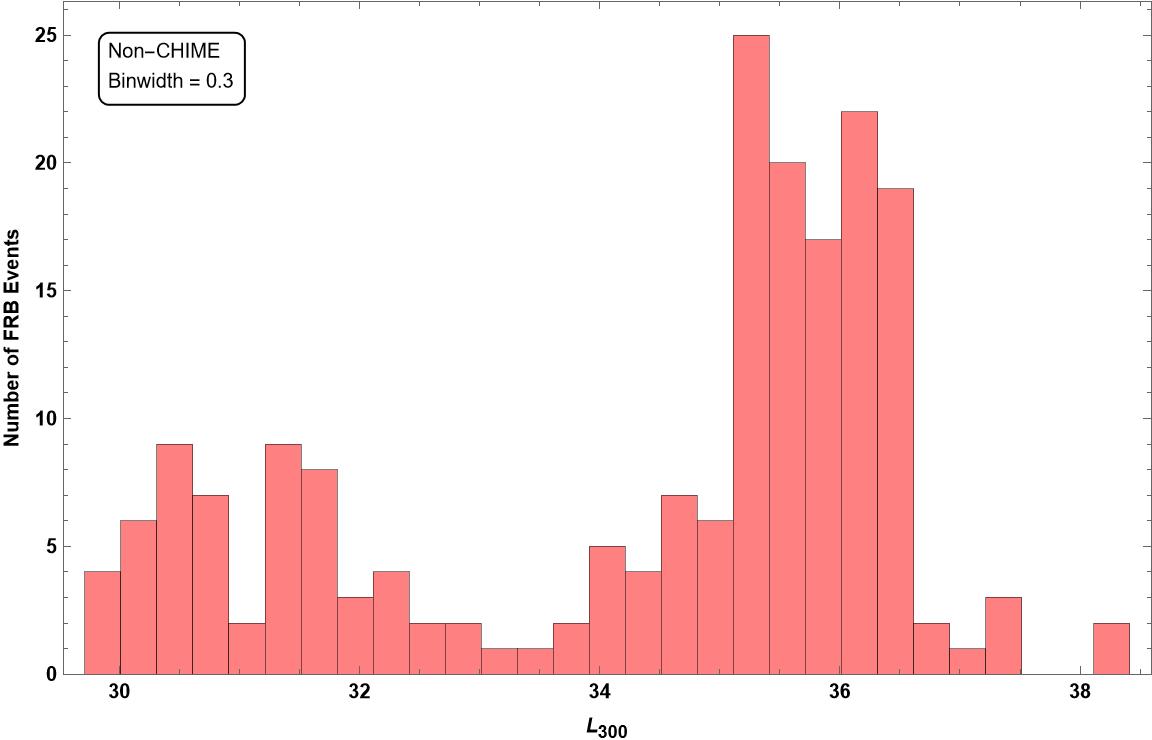}}
    \subfigure[]{\includegraphics[width=0.44\textwidth]{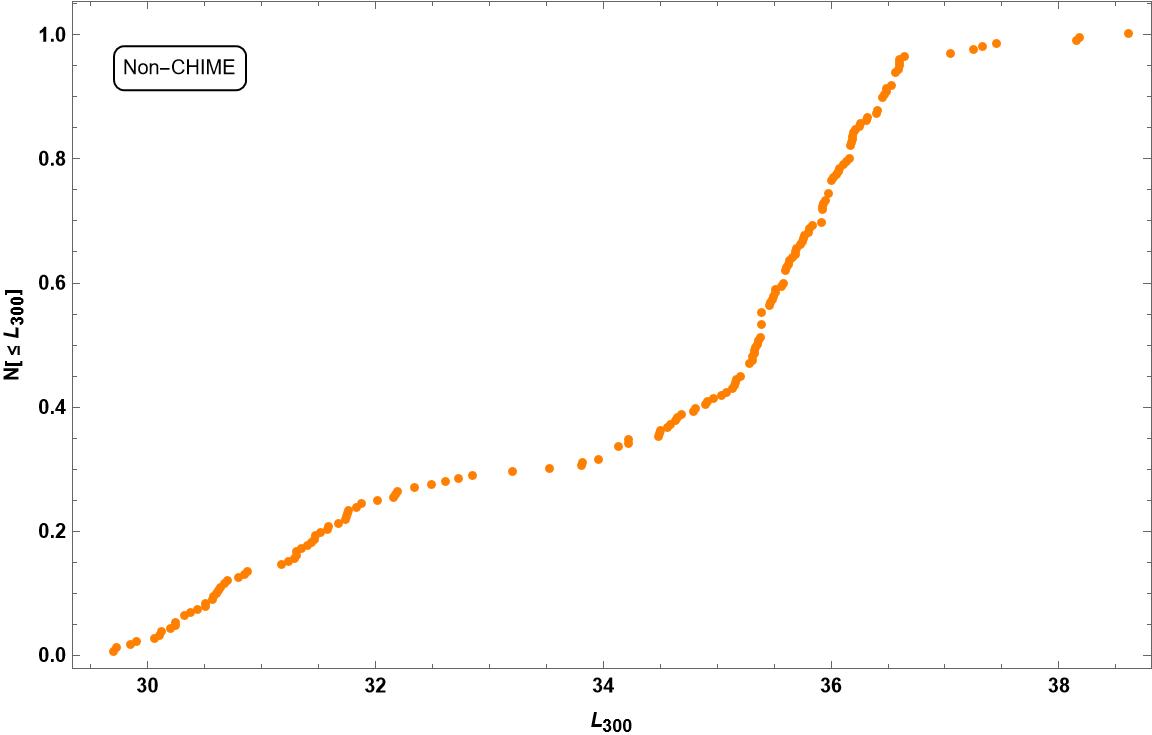}}
    \caption{ (a) A histogram for $L_{300}$ shows a bi-modality with a minimum at $L_{300} \cong 3.98 \times 10^{33}$ erg/sec/Hz. (b) A bin-width independent way of arriving at the same conclusion is seen from the observed $N (\leq L_{300})$ versus  $\log_{10}{(L_{300})}$ in which the slope is close to $\cong 0$ around $L_{300} \cong 3.98 \times 10^{33}$ erg/sec/Hz.}
    \label{fig1}
\end{figure}

The distributions of FRB luminosity density, energy, energy density as well as the ratio of luminosity density to the energy have been studied comprehensively. Figures (\ref{fig1}) to (\ref{fig6}) describe the trends observed for the non-CHIME FRBs. Whenever necessary, the median values of the plotted quantities have been specified in the figures, separately for REPs and NRs.  The luminosity density could be calculated for 194 FRB events based on the available flux density and the estimated redshift data.

The histograms of luminosity density at 300 MHz show bimodal distributions for all five trials of randomly assigned spectral indices to the REPs as well as for the sixth trial. For one such trial corresponding to the random assignment of spectral indices, the histogram is shown in fig.(\ref{fig1}a). The number distribution $n(L_{300})$ exhibits a minimum at $L_{300} \cong 4 \times 10^{33}$ erg/s/Hz. This is also verified in a bin-width independent analysis by considering the slope of,
\begin{equation}
\label{Nl}
    N(\leq L_{300}) \equiv \int^{L_{300}}_0 {n(L)dL}
\end{equation}
as a function of the luminosity density at 300 MHz. It is expected that if $n(L_{300})$ has a minimum, the slope will change from being positive to $0$ (near the location of the minimum) and then positive again. Indeed, as can be seen from fig.(\ref{fig1}b), this slope vanishes around  $ \cong 4 \times 10^{33}$ erg/s/Hz, thus confirming the inference that was based on fig.(\ref{fig1}a). 

The minimum occurs at   $ \cong 4 \times 10^{33}$ erg/s/Hz for all the five trials that use random assignment of $\alpha $ values. For the sixth trial (\textit{i.e.,} $\alpha =1.5 $ case), the minimum is located at $ \cong 5 \times 10^{33}$ erg/s/Hz. When an identical exercise was carried out for the luminosity density at 1 GHz case, the minimum in the distribution was found to be slightly lower, at  $ \cong  10^{33}$ erg/s/Hz. 

Hence, from the above analysis, we conclude that irrespective of the assignment of the spectral index value, the individual FRB events may be classified into two categories - high L, those with $L_{300} > 4 \times 10^{33}$ erg/s/Hz, and a remaining weaker lot, low L,  $L_{300} \lesssim  4 \times 10^{33}$ erg/s/Hz.  To check for the robustness of this classification, a Kolmogorov-Smirnov (K-S) test was performed for the non-CHIME data by considering two of the independent trials on the random assignment of $\alpha $ values to the repeating events, and asking whether both of them were realizations of a given distribution.

For 194 distinct FRB events, the K-S Statistic is 0.0773196 corresponding to the p-value 0.137894. Therefore, the null hypothesis that both of the trials come from the same parent distribution is not rejected at the 5 percent level. It is natural to ask whether the repeating and non-repeating FRBs display marked differences as far as low L and high L classification is concerned.

Fig.(\ref{fig2}) displays the histograms of $L_{300} $ distributions,  distinguishing the FRB events belonging to REPs (represented by lilac color) from that of the one-off FRBs (in light blue), for all the six independent trials. From the median values and the spread, in all six trials,  it is evident that REPs on average fall in the low L category. The histograms of energy $E$ corresponding to the six trials show minima that vary appreciably with individual trials, and hence, the evidence for bi-modality is not very strong.   

Fig.(\ref{fig3}a) and Fig.(\ref{fig3}b) show the distributions of  $E$ for (a) one of the trials on random assignment of $\alpha$ values and for (b) the sixth trial \textit{i.e.,} $\alpha=1.5$ to the REPs, respectively. From the median values of $E$ as well as the spread, one may infer that REPs tend to have lower values of $E$ compared to those of the NRs. The scatter plot given in fig.(\ref{fig4}a) confirms this inference and points to the positive statistical correlation between luminosity density and energy of FRBs. 

On the other hand, fig.(\ref{fig4}b) shows that for a given observed duration $\Delta t_O$, a non-repeater tends to be associated with a somewhat greater lower limit to the energy density, \textit{i.e.} $u$. Since $u \approx E/(c \Delta t)^3 $ the negative correlation seen in this figure between $u$ and $\Delta t_O$ is expected. From the histograms of figures (\ref{fig5}a) and (\ref{fig5}b) (and the corresponding median values), it is quite clear that NRs tend to have larger values of $u$, which is consistent with the corresponding dimensionless $X5$ values.

The histograms presented in fig.(\ref{fig6}) show that, for all the five trials, distributions of the ratio of $L_{300}$ to the energy $E$ as well as the corresponding median values point to REPs tending to have lower values of this dimensionless quantity compared to that of the NRs. However, the difference is very marginal for the sixth trial (\textit{i.e.,} $\alpha=1.5$ for all the REPs). The interesting point is that since $L_{300}/E$ involves the ratio $S_{\nu_O}/F_{\nu_O}$, this analysis can be carried over to the CHIME FRBs as well.

\subsection{CHIME}
Given that many of the FRBs detected by CHIME are from off the center of the latter's beam pattern, one can only set lower limits to the FRB's actual flux density and fluence. However, for a CHIME FRB, it is reasonable to assume that the ratio of its measured flux density to fluence is close to the true ratio had the FRB been detected in the direction coinciding with the beam center of the telescope. So, to compare the results we have obtained in the case of non-CHIME FRBs with the CHIME ones, we have introduced a new technique whereby we use the quantities $L_{300}/E$, $X1$, $X2$, $X3$, $X4$, $X6$ and $X7$  that involve the ratio $S_{\nu_O}/F_{\nu_O}$ (or its reciprocal) in our analysis for both these data sets.

The other advantage of this technique is that the numerical values of each of the parameters (eqs. (25)-(35)) as well as $L_{300}/E$ lie in the same range for both CHIME and non-CHIME FRBs leading thereby to a meaningful comparison of CHIME and non-CHIME FRBS. This feature also supports our novel use of considering the ratio of flux density to fluence in defining the constructed parameters. For instance, the histograms of figs.\ref{fig6} (a)-(f) demonstrate that distributions of $L_{300}/E$ and the corresponding median values of non-CHIME FRBs show the same trends as in the case of CHIME (figs.\ref{fig7} (a)-(f)), entailing an internal consistency of our new method. The bi-modality in the distribution of $L_{300}/E $ is displayed in both CHIME and non-CHIME FRBs, albeit the evidence being weaker when all the repeaters are assigned $\alpha = 1.5$. This encourages us to claim that even CHIME FRBs would have shown a bi-modal distribution in $L_{300} $ had the measured values of the flux density for each of the FRBs were available.

From fig.(\ref{fig8}) to fig.(\ref{fig12}), the distributions of the dimensionless quantities have been shown for both non-CHIME (left-hand panel) and CHIME FRBs (right-hand panel) for comparisons. There is an overall agreement in the FRBs' trends in the two distinct data sets.  Figs. (\ref{fig13}a) and (\ref{fig13}b), although not dimensionless but involves the ratio $F_{\nu_O}/S_{\nu_O}$, show that the ratio of the lower limit to the energy density to the luminosity density tends to be larger for REPs in comparison to that of the NRs for non-CHIME FRBs and marginally larger for the CHIME ones. The histogram of $k_B T_b/E$ for the CHIME FRBs fig.(\ref{fig12}b) suggests that the NRs tend to have higher brightness temperatures than their repeating counterparts, although this trend is marginal for the non-CHIME FRBs fig.(\ref{fig12}a). 

Using both supervised~\cite{38} and unsupervised~\cite{50} machine learning methods, a recent study aimed to look for potential repeating FRBs among NRs in the CHIME first catalog. According to their findings, few of the earlier classified NRs are predicted to be potential REPs suggesting a search for recurring radio transients for these FRBs. The authors classify those potential REPs as strong which have been identified in both of the papers~(\cite{38},\cite{50}). In the scatter plots of our present study, we represent the strong REP candidates with a magenta asterisk, while those identified via an unsupervised (supervised) machine learning as type I (type II) REPs with a blue (red) asterisk. In our analysis, from figs. (\ref{fig8}b) to (\ref{fig11}b), it is very interesting to note that the majority of these potential REPs indeed follow the patterns that are common to the standard REPs.

Figs. (\ref{fig14}a) and (\ref{fig14}b) are for the CHIME bursts alone corresponding to a trial involving random assignment of $\alpha  $ values to the REPs. The distributions of  $X2$ and $X3$ as well as their median values reflect the tendency of NRs to have higher values than those of the REPs. Marginal bi-modalities are seen in the distributions of REPs entailing a hint of two distinct modes of radio-emission from the REPs. It may be pointed out, that these trends survive even for the sixth trial (i.e. $\alpha =1.5 $ for all the REPs).  More future detections of REPs are required to strengthen this surmise. The histogram of the flux-density to fluence for CHIME FRBs is shown in (\ref{fig14}c). Although the REPs and NRs do show a difference in the distribution, there is no bi-modality in the distribution as such.

The histograms as well as scatter diagrams of $X7$ have been plotted in (\ref{fig15}a), (\ref{fig15}b), (\ref{fig16}a)  and (\ref{fig16}b). The REPs tend to have a smaller median value of $X7$ compared to that of NRs. It is important to emphasize here that the parameter $X7$ involves the true brightness temperature and energy density even though the size of the emission region is not known to us.

\section{Simple Theoretical Considerations: Pulsar/Magnetar Glitches, Goldreich-Julian Charge Density and FRBs}

Since many pulsars exhibit rapid spin-up at times, with $\Delta \omega/\omega \cong 10^{-11} - 10^{-5}$, it is natural to expect theoretical models linking FRB activities to sudden glitches of magnetized neutron stars \cite {51,52}.
The recent detection of an anti-glitch $|\Delta \omega /\omega | \cong 5.8 \times 10^{-6}$ from the magnetar SGR 1935+2154, followed by three radio transients from it akin to FRBs \cite{53}, has given boost to such models \cite {51,52,55}. Earlier, another magnetar 1E 2259+586 had also exhibited an anti-glitch \cite{54}. It has also been reported that  SGR 1935+2154 displayed a very large glitch  $\Delta \omega /2\pi  \cong +  19.8 \times 10^{-6}$ Hz before the FRB 200428 event \cite{56}. Large glitches can also significantly impact the magnetosphere and inner gap of pulsars, causing pulsars below the death line to get activated \cite{57,58}.

We have considered in this paper a very simple physical scenario connecting FRBs to magnetar glitches.
Assuming a spinning neutron star (NS)  of radius $R$, angular velocity $\vec {\omega} = \omega \hat{k} $ and magnetic field strength $B_p$ at the poles,  the exterior magnetic field distribution is that of  a magnetic dipole moment $\vec{m} $ making an angle $\xi $ with respect to  the spin axis, the time-dependent magnetic field at $\vec{r} $ within the light-cylinder  is given by,
\begin{equation}
    \vec{B} (\vec{r},t) = \frac{1}{r^5} [3 \vec{r} \ (\vec{r}.\vec{m} (t)) - r^2 \vec{m} (t) ]
\end{equation}
where,
\begin{equation}
    \vec{m} (t)= \frac{B_p R^3} {2} [\sin {\xi}\ (\cos{\phi (t)}\ \hat {i} + \sin{\phi (t)} \ \hat {j} \ ) + \hat{k} \ \cos{\xi} ]
\end{equation}
with,
\begin{equation}
    \phi (t) = \phi - \int^t_0 {\Omega (t^\prime) dt^\prime} \  
\end{equation}
and for $t > t_{gl}$,
\begin{equation}
    \Omega(t) \cong \omega + \Delta \omega\  [1- Q \ (1- e^{-\frac{(t - t_{gl})} {\tau}})]
\end{equation}
taking into account the possible occurrence of a glitch at time $t_{gl} $ (see e.g. \cite{59}).
 For $t <  t_{gl}$,  the angular speed of the NS is simply $\omega $ (We have ignored the slow and monotonic decrease in $\omega$ due to the magnetic braking, since the time interval over which FRB activities take place is much shorter in comparison.). 

In eq.(40), $\Delta \omega $ is the jump in the angular speed, $Q$ is the healing parameter while $\tau$ is the relaxation time for the angular speed to return to its original value that depends on the moments of inertia of the superfluid core and the full NS as well as the frictional coupling between the superfluid neutrons with the NS crust. For pulsars like Crab and Vela, the parameters $Q$ and $\tau$ are usually less than unity and several tens of days, respectively  \cite{59}.

Assuming a spherical polar coordinate system with the origin at the center of the NS, the Goldreich-Julian (GJ) charge density within the light cylinder radius $R_{lc} (t) \equiv (r \sin{\theta})_{lc}= c/\Omega(t)$ corresponding to an oblique rotator, like  the above, is given by {\cite{60,61}},

\begin{strip}
\begin{equation}
 \rho_{GJ} (\vec {r},t) = - \frac {B_p R^3 \ \Omega(t)}  {4\pi c \  r^3} \bigg ( \frac {3\cos{\theta}[\cos{\xi} \cos{\theta} + \sin{\xi} \sin{\theta} \cos {\phi (t)}] - \cos{\xi}}{1- \Omega^2 (t) r^2 \sin^2{\theta}/c^2} \bigg )  \ \ ,
\end{equation}
\end{strip}
after making use of eq.(38) as well as incorporating a time-dependent angular speed $\Omega (t)$ to include the effect of a glitch/anti-glitch. 
Eq.(41)  is a straightforward generalization of the expression derived by Melrose and Yuen \cite{61} for an oblique rotator with a constant spin angular speed.
%%%%%
Before a glitch takes place, the GJ density below  the light-cylinder $r \approx (R_{lc} - \Delta r)/\sin{\theta}$ with $0 < \Delta r \ll R_{lc} $, can be obtained from eq.(41),
\begin{equation*}
  (\rho_{GJ})_{\mbox{\tiny{pre-glitch}}} \bigg(\frac{R_{lc} - \Delta r}{\sin{\theta}},\theta, \phi , t \bigg)  = - \frac {B_p R^3 \ \omega}  {4\pi c \ (R_{lc} - \Delta r )^3}   \times 
\end{equation*}
\begin{equation}
    \qquad \bigg ( \frac {f(\theta, \phi , \xi, \omega , t)}{1- \omega^2  (R_{lc} - \Delta r )^2 \sin^2{\theta}/c^2} \bigg )  \ \ ,
\end{equation}
where the factor $f$ containing the angles is given by,
\begin{equation*}
f(\theta, \phi , \xi, \omega, t)= \sin^3{\theta} \ [3 \cos{\theta}(\cos{\xi} \cos{\theta} + 
\end{equation*}
\begin{equation}
  \qquad + \sin{\xi} \sin{\theta} \cos {(\phi - \omega t)}) - \cos{\xi}]  \ .
\end{equation}
In  eq.(42), the presence of $R_{lc} - \Delta r$, however tiny $\Delta r $ is, ensures that the denominator does not vanish. 

If FRB phenomena is caused by a glitch/anti-glitch the associated time scale, on the other hand, is expected to be much less compared to $\tau$.  Hence, post glitch, the angular speed  given by eq.(40)  during $ t_{gl} < t \ll t_{gl} + \tau $ can be approximated to,
\begin{equation}
\Omega (t)  \cong  \omega + \Delta \omega \ [1 - Q \ \frac {t- t_{gl}} {\tau}] = \omega + \delta \omega (t)\ \ ,
\end{equation}
where $\delta \omega (t) \equiv  \Delta \omega \ [1 - Q \ \frac {t- t_{gl}} {\tau} ]$. Eq.(42) leads to $\phi(t)$ (eq.(39)) having the form,
\begin{equation}
    \phi(t) \cong \phi - (\omega t +  (t-t_{gl}) \delta \omega (t)) \ \ .
\end{equation}

Therefore, making use of eqs.(41), (43) and (44), the  post-glitch GJ density  of charged particles near  $r \sim (R_{lc} - \Delta r)/\sin{\theta}$ is given by,
\begin{equation*}
(\rho_{GJ})_{\mbox {\tiny{post-glitch}} } \bigg(\frac{R_{lc} - \Delta r}{\sin{\theta}},\theta, \phi , t \bigg) \cong - \frac {B_p R^3 \ \omega (1+ \delta \omega (t)/\omega )}  {4\pi c\  R^3_{lc} \ (1 - \Delta r/R_{lc} )^3}   \times 
%(\frac{R_{lc} - \Delta r}{\sin{\theta},\theta,t) \cong \bigg (\frac{3\cos{\theta}\ [ \cos{\theta} + \tan{\xi} \sin{\theta} \cos {\phi (t)}] - 1}{3\cos{\theta}\ [ \cos{\theta} + \tan{\xi} \sin{\theta} \cos {\omega t}] - 1} \bigg ) \times 
\end{equation*}
\begin{equation}
\times \bigg ( \frac{f(\theta, \phi , \xi, \omega + \delta \omega (t), t) }  {1 - R^2_{lc} \omega^2 (1+ \delta \omega (t)/\omega )^2 (1 - \Delta r/R_{lc} )^2  /c^2)}\bigg )\ \ .
\end{equation}

Coherent radio emission is expected to occur along the open magnetic field lines in the regions just beyond the light-cylinder (LC).
When a glitch takes place, the size of the LC  shrinks suddenly for a brief time interval,
\begin{align}
    \delta R_{lc} &=   - R_{lc} \frac {\delta \omega (t)} {\omega}  + \mathcal{O} \bigg (\delta \omega^2/\omega^2 \bigg ) \\
     &\cong - 2.4 \times 10^5  \ \bigg (\frac {P} {5 \ s} \bigg ) \ \frac {\delta \omega (t)} {\omega}\ \mbox{km} \ \ ,
\end{align}
assuming $\Delta \omega /\omega  \ll 1$, so that charge particles in the immediate vicinity of the pre-glitch LC find themselves outside where open field lines are now present. In the above equation, $P$ is the spin period of the NS. It is interesting to note from eq.(48) that $|\delta R_{lc}| /c  \sim 10^{-3} \ s$, so relevant for FRBs,  when $\delta \omega /\omega \sim 10^{-3}$. 

Taking the volume of emission region to be  $\Delta V \sim |\delta R_{lc} |^3 \approx R^3_{lc} (\delta \omega /\omega)^3 $ and using eq.(46),  the estimated number of charged particles, $ \delta N_e $, moving along the open field lines  is given by,
%\begin{strip}
\begin{equation*}
    \delta N_e = \frac{(\rho_{GJ})_{\mbox {\tiny{post-glitch}} }} {e} \Delta V \cong 
     - \frac {B_p R^3 \ \omega }  {8\pi \ e \ c\   \ }  \bigg (1+ \frac{\delta \omega (t)}{\omega} \bigg ) \ \times 
%(\frac{R_{lc} - \Delta r}{\sin{\theta},\theta,t) \cong \bigg (\frac{3\cos{\theta}\ [ \cos{\theta} + \tan{\xi} \sin{\theta} \cos {\phi (t)}] - 1}{3\cos{\theta}\ [ \cos{\theta} + \tan{\xi} \sin{\theta} \cos {\omega t}] - 1} \bigg ) \times 
\end{equation*}
\begin{equation}
\times \ \bigg (\frac{\delta \omega (t)}{\omega}\bigg )^3  \bigg (1 + \frac{3 \Delta r}{R_{lc}} \bigg )  \bigg ( \frac{f(\theta, \phi ,  \xi, \omega + \delta \omega (t), t) }  {\Delta r /R_{lc} - \delta \omega (t)/\omega} \bigg )\ \ .
\end{equation}
In the above equation, the term $ \Delta r /R_{lc} - \delta \omega (t)/\omega = \Delta r /R_{lc} - |\delta R_{lc} | /R_{lc}$ can be very close to zero, since $\Delta r$ and $|\delta R_{lc} | $ are both $\ll R_{lc}$. Hence, 
introducing a dimensionless parameter $\eta (t)$ so that,
\begin{equation}
  \frac {\Delta r} {R_{lc}} \equiv \eta \bigg \vert  \frac {\delta R_{lc}}{R_{lc} } \bigg \vert \Rightarrow \frac {\Delta r} {R_{lc}} - \frac {\delta \omega (t)} {\omega} = (\eta -1)  \frac {\delta \omega (t)} {\omega} \ \ ,
\end{equation}
we may recast eq.(49) as,
%\begin{strip}
\begin{equation*}
    \delta N_e = 6 \times 10^{29} \ \frac{(\delta \omega/\omega)^2}{\eta-1} \bigg(1+\frac{\delta \omega}{\omega}\bigg) \bigg(1+\frac{3 \eta \delta \omega}{\omega}\bigg)\  f \ \times
\end{equation*}
%\end{strip}
\begin{equation}
    \qquad \times \bigg( \frac{B_P}{10^{14} \ \mbox{Gauss}} \bigg) \bigg( \frac{R}{12 \ \mbox{km}} \bigg)^{3} \bigg( \frac{P}{5\ \mbox{s}}\bigg)^{-1} \ \ .
\end{equation}

FRB brightness temperatures being rather high, we may invoke curvature radiation due to $N_{\mbox{\tiny{bunch}}} $ of charge particles moving coherently along the open field lines just outside $R_{lc}$ \cite{22,23}. Assuming $N_{\mbox{\tiny{bunch}}} \sim  \delta  N_e $, the radio-luminosity ensuing from curvature radiation is  given by,
%\end{strip}
%$$ \delta N_e \approx l^3_{em} \delta n_e \lesssim (c \Delta t)^3 \delta n_e $$
%\begin{align}
%    \delta N_e &\approx l^3_{em} \delta n_e \lesssim (c \Delta t)^3 \delta n_e  \\
 %    &\cong (1.9 \times 10^{26}) \bigg (\frac{B_z} {10^{14} \ \mbox{Gauss}} \bigg ) \bigg (\frac{P}{1 \ s} \bigg )^{-1} \bigg %%(\frac{\Delta t}{1 \ ms} \bigg )^3 \ \ .
%\end{align}

\begin{align}
L &= \frac {2 \gamma^4 N^2_{\mbox{\tiny{bunch}}} e^2  c}{3 R^2_{curv}} \\
 \qquad &\approx 3.2 \times  10^{43}\ \bigg (\frac {N_{\mbox{\tiny{bunch}}}} { 10^{29}} \bigg )^2 \bigg (\frac {\gamma} {10^2} \bigg )^4 \bigg (\frac {R_{curv}} {120\ \mbox{km}} \bigg )^{-2} \ \mbox{erg} \ \mbox{s}^{-1}
\end{align}
 at a  characteristic radio frequency,
\begin{equation}
\nu_{curv}= \frac {3 \gamma^3 c} {4 \pi R_{curv}} \approx 590  \ \bigg (\frac {\gamma} {10^2} \bigg )^3 \bigg (\frac {R_{curv}} {120\ \mbox{km}} \bigg )^{-1} \ \mbox{MHz} 
\end{equation}
where $R_{curv} $ is the curvature radius of the field lines that is expected to be $\sim 10\ R$ \cite{62}.

From the results discussed in $\S 4.1$ pertaining to the bi-modality in the luminosity density distribution, the median values of $L_{300}$ for repeating and non-repeating FRBs are $\cong  10^{32}$ erg/Hz/s and $\cong  10^{35}$ erg/Hz/s, respectively. These median values correspond to  median radio-luminosities $\cong 2 \times 10^{40}$ erg/s and $\cong 2 \times 10^{43}$ erg/s, respectively,  if one takes the spectral index $\alpha=1.5 $.  This bi-modal feature can be explained, by making use of eqs.(51) and (53),  if one considers the polar magnetic field to be  $B_p \cong 10^{14}$ Gauss,   $\eta -1 \cong  8 \times 10^{-8}$ and $\delta \omega/\omega \cong 1.83 \times 10^{-5}$ for the weaker category of FRBs. For the stronger category, the magnetic field needs to be one order of magnitude higher, i.e. $B_p \cong 10^{15}  $ Gauss and $\delta \omega/\omega \cong 3.25  \times 10^{-5}$, keeping the value of $\eta -1 $ the same. On the other hand, if  $B_p=10^{14} $ Gauss is maintained for  the high luminosity category of FRBs, the median value of  $\delta \omega/\omega $ needs to go up to $\cong  10^{-4}$.

%It is very interesting to note that coherent radio emission due to a magnetic field-induced acceleration of bunched charged particles to explain FRB phenomena that we had considered in an earlier study of ours requires bunching of $\sim 10^{25}$ electrons (positrons)\cite{23}.

%As we can see from the above the luminosity is very sensitive to the spin period of the neutron star. 

\section{Discussions and Conclusion}
 
We have found several interesting patterns based on a study of the distributions of several parameters associated with non-CHIME and CHIME FRBs. Our robust conclusion is that non-CHIME FRBs come in two distinct categories based on their $L_{300}$ distribution: low and high luminosity density FRBs. The category with $L_{300} \lesssim 4 \times 10^{33}$ erg/Hz/s  consists largely of repeating FRBs, while most of the non-repeaters have $L_{300} \gtrsim 4 \times 10^{33}$ erg/Hz/s.  Less significantly, the FRB radio energy $E$  as well as a lower limit to the FRB energy density (i.e. $u$) show weak bi-modal distributions. 

 Introducing a new technique wherein several quantities have been computed that utilize the ratio of the observed fluence to flux density (or, its reciprocal). This exercise was undertaken since only lower limits to the flux density and fluence are known for the CHIME FRBs and therefore, using this technique,  both CHIME and non-CHIME FRBs could be studied. Distributions of these parameters as well as their median values suggest the presence of significant trends common to both the data sets. 

For instance, the distribution of the dimensionless quantity $L_{300}/E$ computed for the CHIME FRBs exhibits a bi-modality that is very similar to that of the non-CHIME FRBs, with a majority of the non-repeaters falling in the category of higher values of $L_{300}/E$. This strengthens our argument favoring the existence of two distinct categories of FRB events wherein repeaters tend to be associated with lower values of radio luminosity density events.  

The other robust conclusion in this regard is that the values of the quantities constructed using $S_{\nu_O}/ F_{\nu_O}$ are almost identical and lie in the same ranges whether one is considering CHIME or non-CHIME FRBs.
This universality is also reflected in the distribution patterns of these quantities, buttressing the validity of studying quantities that involve $S_{\nu_O}/ F_{\nu_O}$, in the case of CHIME FRBs.   The observed trends pertaining to repeaters and non-repeaters are strikingly similar for both  CHIME as well as non-CHIME FRBs, particularly when one assigns randomly the spectral index $\alpha $ values to the repeaters. However, when $\alpha$ is set to 1.5 for all the repeaters, several of the trends become marginal. This draws one's attention to the importance of  $\alpha $ value determination for the FRBs. The study of fast radio transients using SKA in the future may throw more light on this crucial aspect \cite{63}.

We have also discussed a simple theoretical model to explain the observed bi-modality in the FRB luminosity. The model hinges on the effect of magnetar glitches on the Goldreich-Julian charge density as well as on the abrupt decrease in the light-cylinder radius. The bi-modality emerges if one considers for the stronger FRB category a polar magnetic field of $\sim 10^{15} $ Gauss and $\delta \omega /\omega \sim 3.25 \times 10^{-5}$ as against the values $\sim 10^{14} $ Gauss and $ \sim 1.8 \times 10^{-5} $, respectively, for the weaker category. Of course, with a larger value of glitch, $\delta \omega /\omega \sim  10^{-4}$, the higher luminosity category FRBs can ensue from magnetars with $ B_p \cong 10^{14}$ Gauss.

Based on the studies of Crab and Vela pulsars, it has also been suggested that strong glitches can lead to precession of the spin axis of compact stars \cite {64}.  Precession may play an important role in explaining the periodic occurrences of some of the repeaters. Similarly, repeaters and non-repeaters could have distinct origins as a class of non-repeating FRBs may be caused by resonant conversion of high frequency gravitational waves into radio by means of the Gertsenshtein–Zel'dovich effect \cite{65}.

\section*{Acknowledgement}
N.S.  thanks Dr. Shriharsh Tendulkar and Dr. Emily  Petroff for their helpful comments on the FRB data catalog. PDG is indebted to  Late Dr. N. Rathnasree for several discussions on pulsar emission models.  We also thank Ketan Sand, for his valuable comments on FRB data, and Pankaj Pawar, for some of the preliminary data analysis work.

%%use \balance somewhere in the left column of the last page to balance the two columns in the end page

%%References section

%\bibitem{latexcompanion} 
%Michel Goossens, Frank Mittelbach, and Alexander Samarin. 
%\newblock {\em The \LaTeX\ Companion}. 
%Addison-Wesley, Reading, Massachusetts, 1993.

%\end{thebibliography}

%******************************************************************************************
%  Figures
%******************************************************************************************

%\twocolumn[{

\begin{figure}[ht]
    \centering
    \subfigure[]{\includegraphics[width=0.44\textwidth]{ 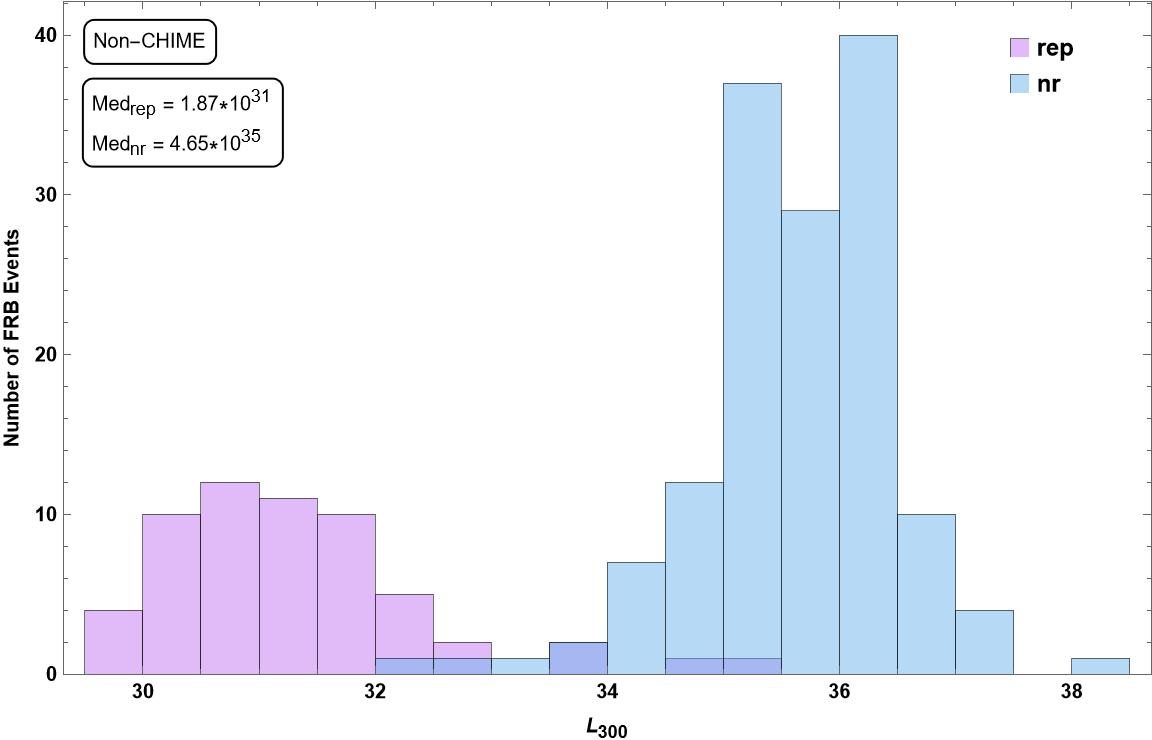}}  
    % \end{figure}
    % \begin{figure}[ht]
    % \addtocounter{subfigure}{0}
    \subfigure[]{\includegraphics[width=0.44\textwidth]{ 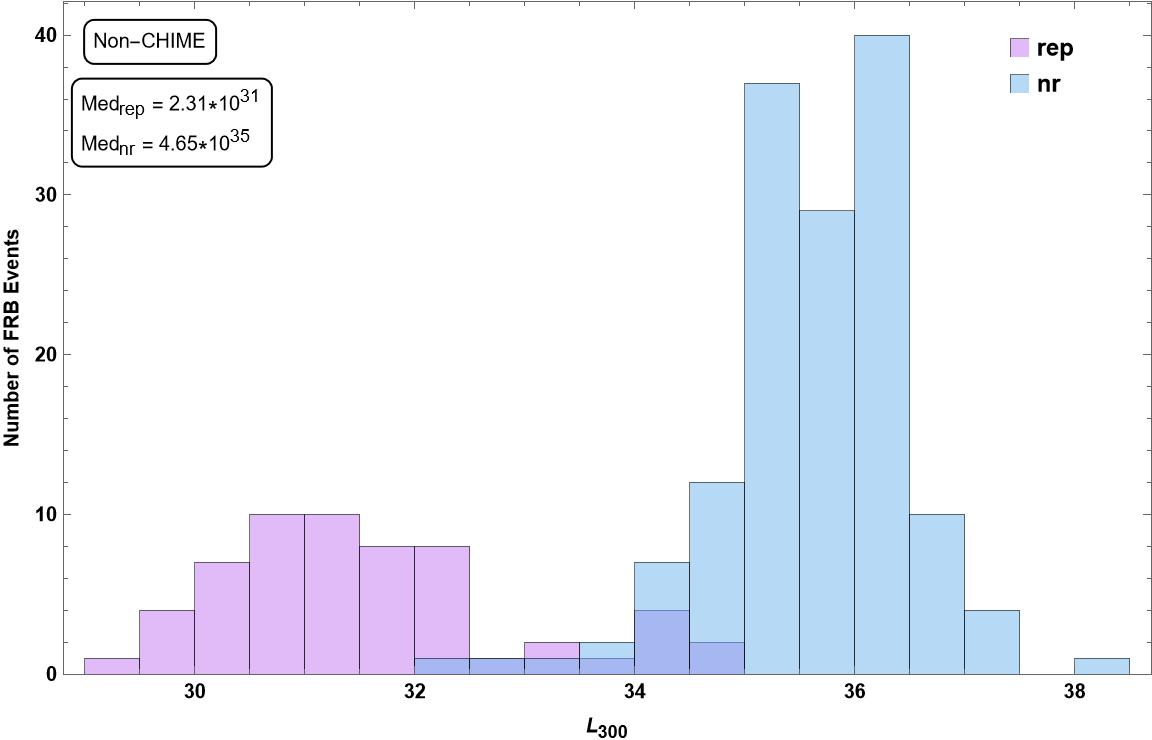}}
    %\vspace{0.1cm}
    \subfigure[]{\includegraphics[width=0.44\textwidth]{ 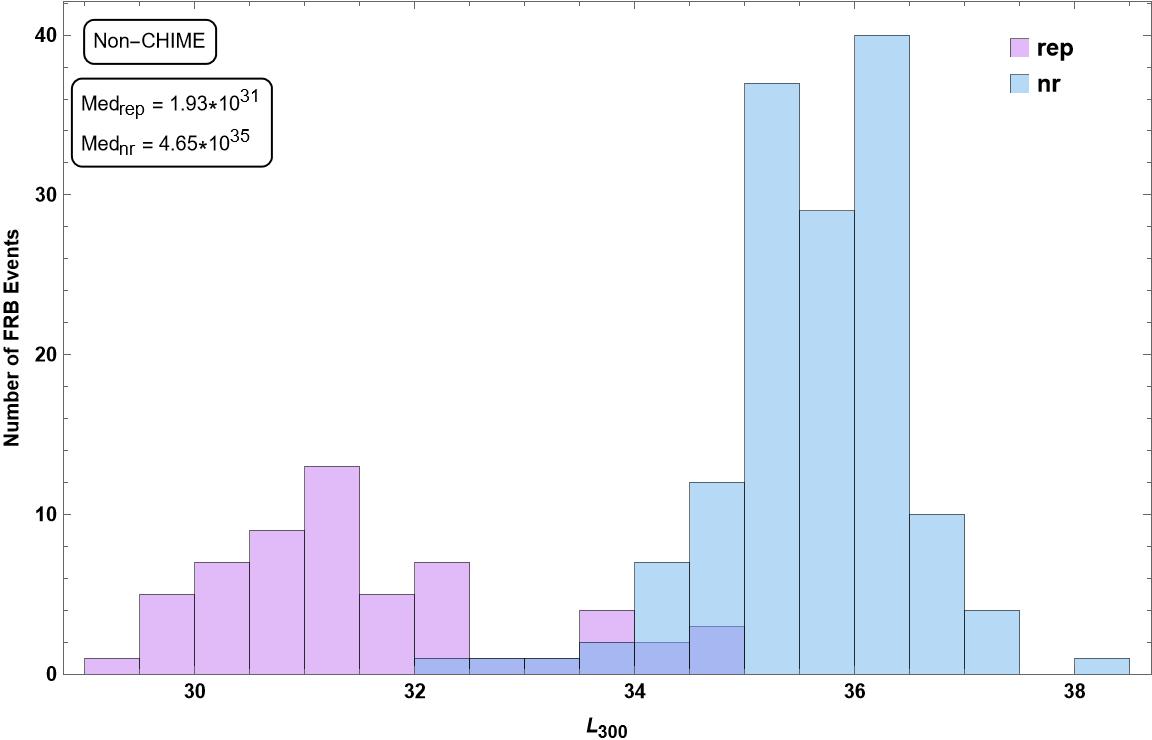}}
    \end{figure}
    \begin{figure}[ht]
    \addtocounter{subfigure}{0}
    \subfigure[]{\includegraphics[width=0.44\textwidth]{ 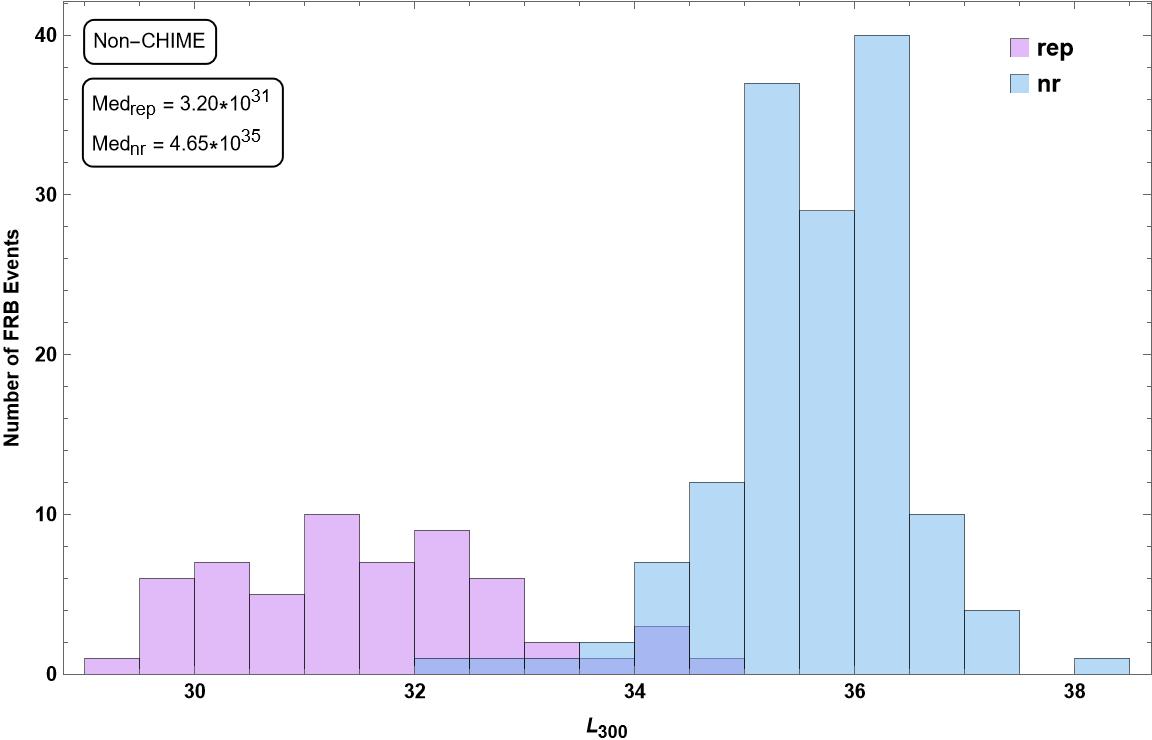}}
    %\vspace{0.1cm}
    \subfigure[]{\includegraphics[width=0.44\textwidth]{ 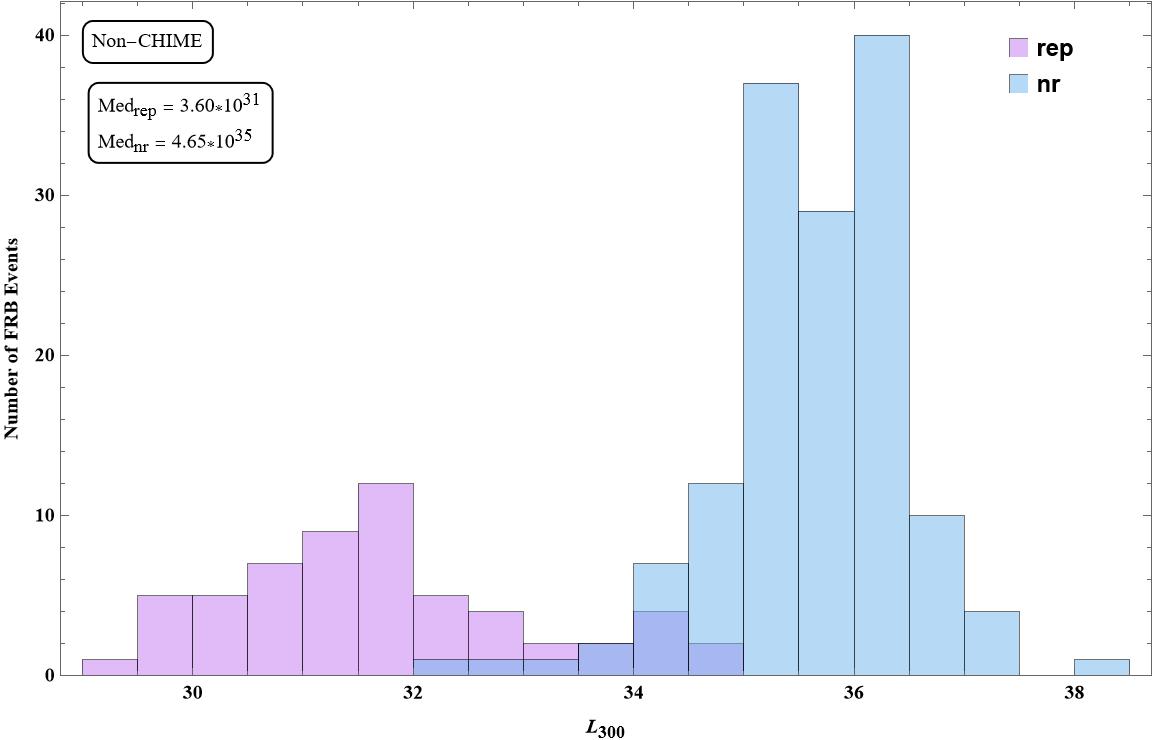}}
    \subfigure[]{\includegraphics[width=0.44\textwidth]{ 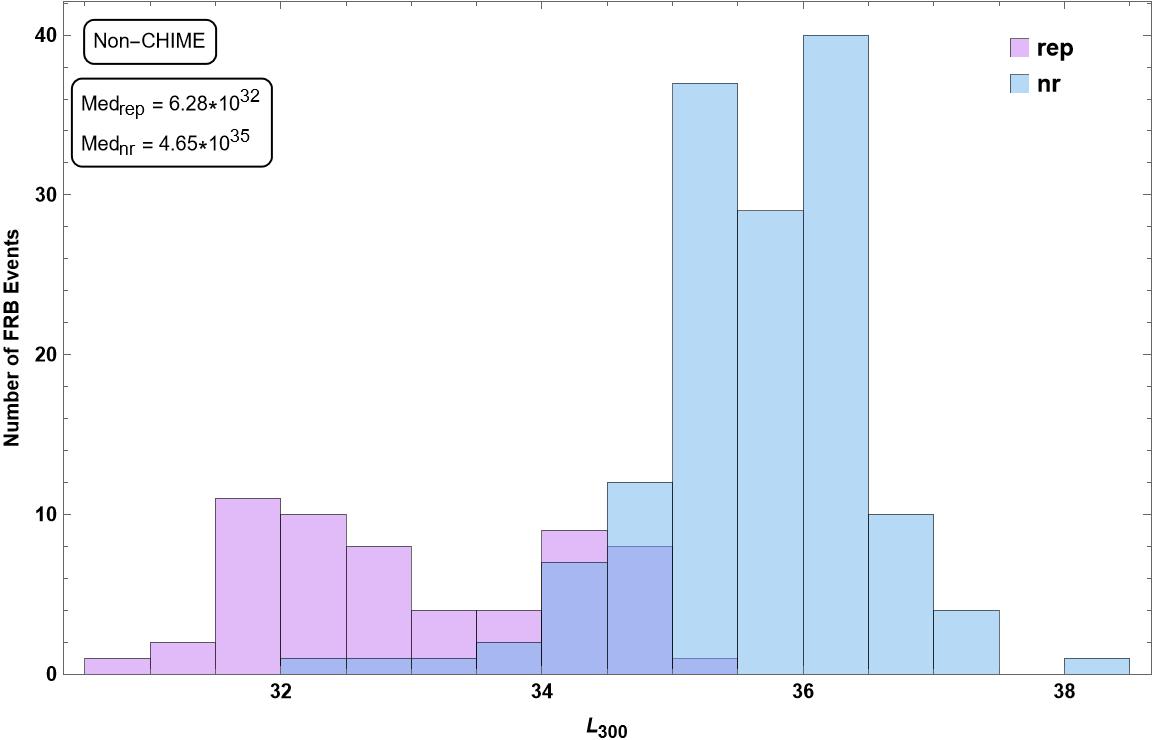}}
    \caption{ Histograms of $L_{300} $ of repeaters and non-repeaters. The first five figures correspond to five different sets in each of which the repeaters are assigned six randomly generated spectral index values. The last histogram is for the case when all the repeaters are assigned  $\alpha$=1.5.}
    \label{fig2}
\end{figure}

\begin{figure}[!ht]
    \centering
    \subfigure[]{\includegraphics[width=0.44\textwidth]{ 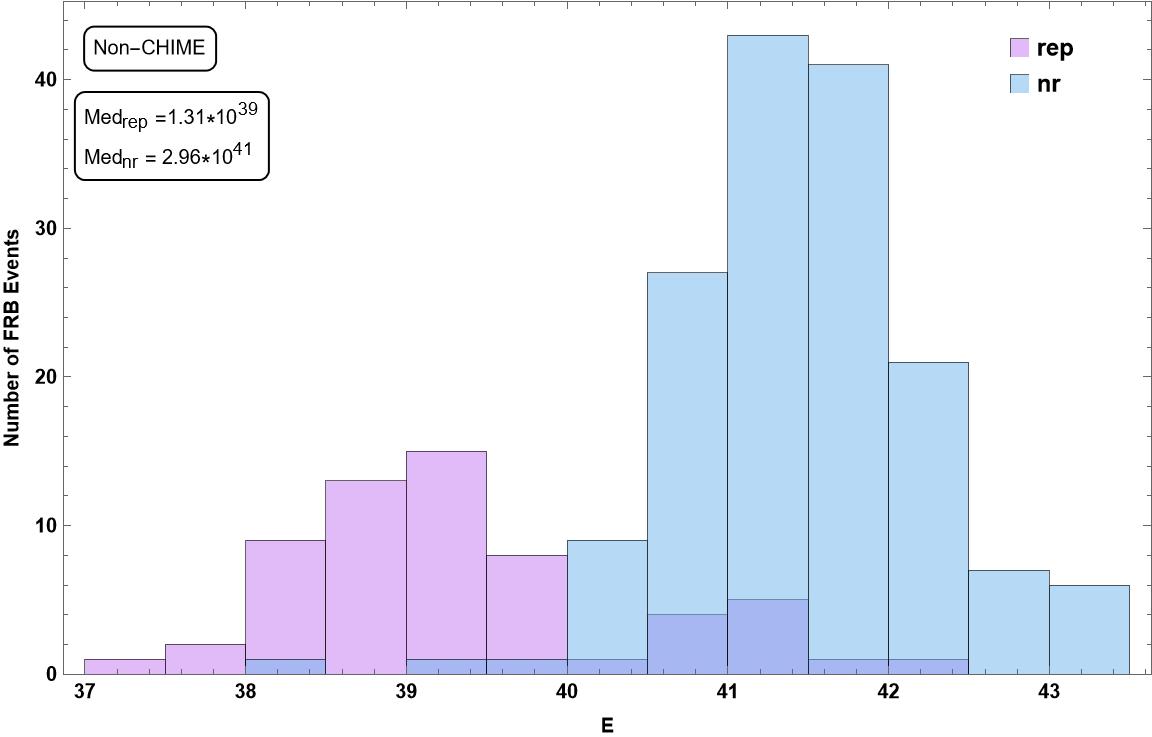}}
    \subfigure[]{\includegraphics[width=0.44\textwidth]{ 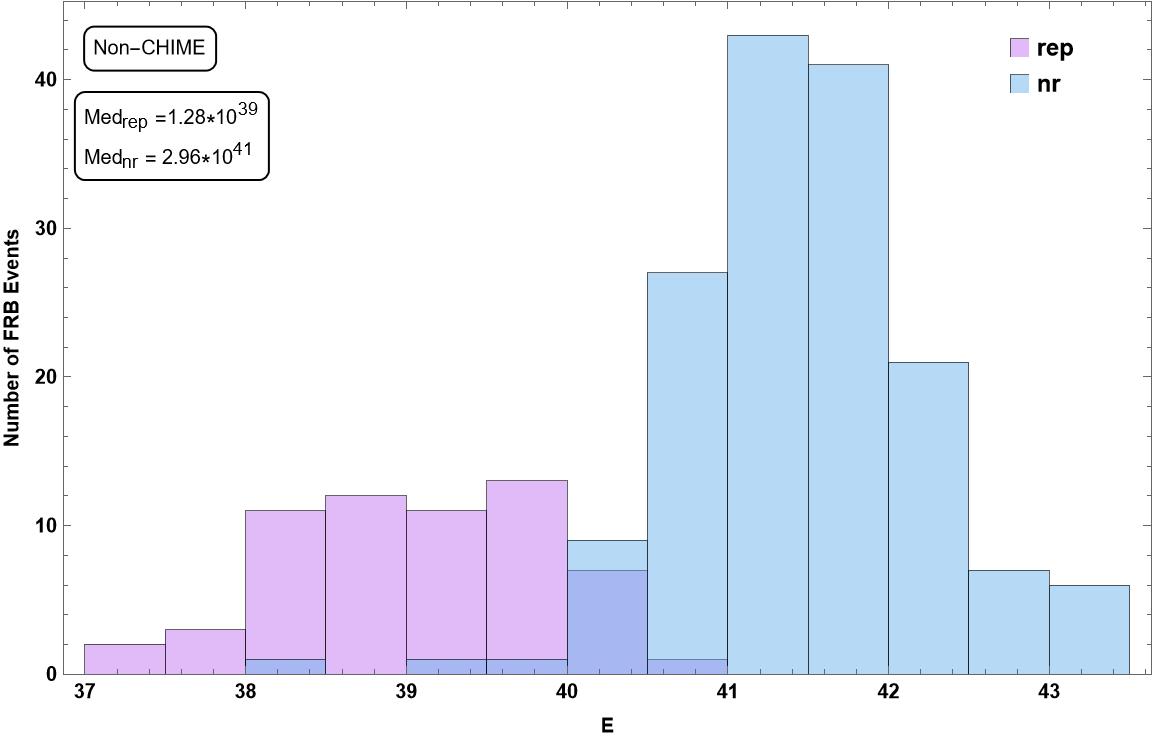}}
    \caption{ (a) A histogram of energy $E$ with repeaters assigned randomly a set of spectral indices.  (b) A histogram of $E$  when $\alpha $ is set to be 1.5 for all the repeaters.}
    \label{fig3}
\end{figure}

\begin{figure}[!ht]
    \centering
    \subfigure[]{\includegraphics[width=0.44\textwidth]{ 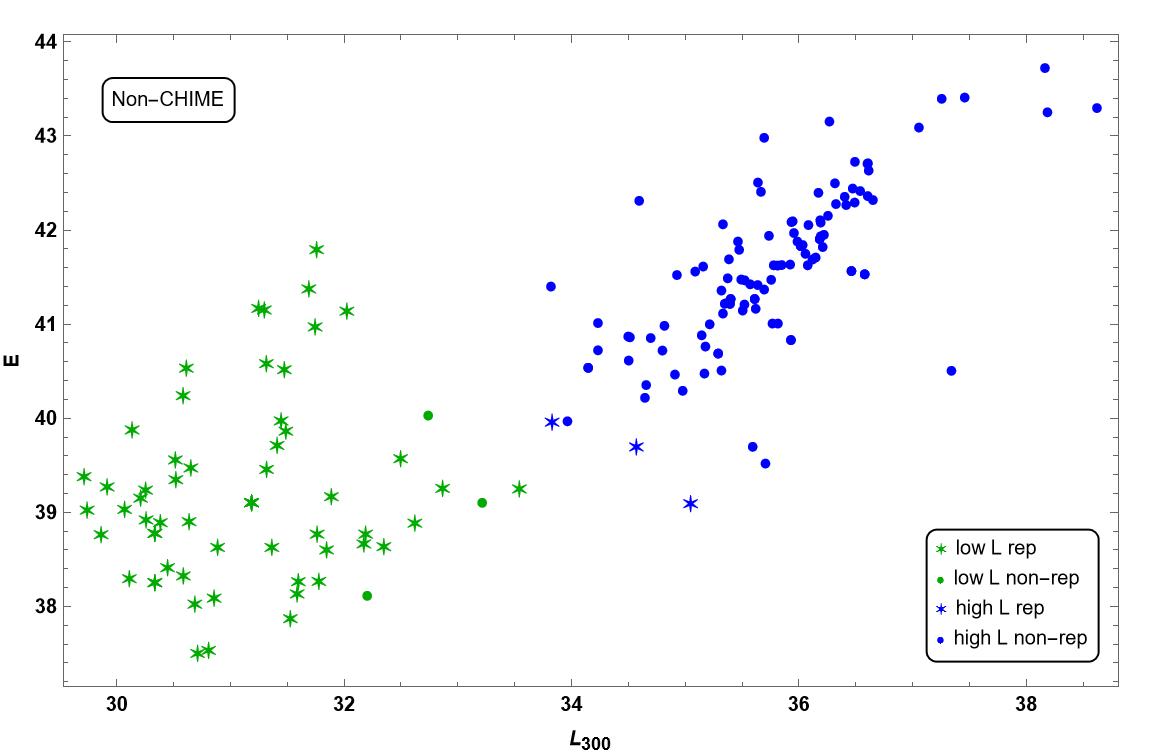}}
    \end{figure}
    \begin{figure}[ht]
    \addtocounter{subfigure}{0}
    \subfigure[]{\includegraphics[width=0.44\textwidth]{ 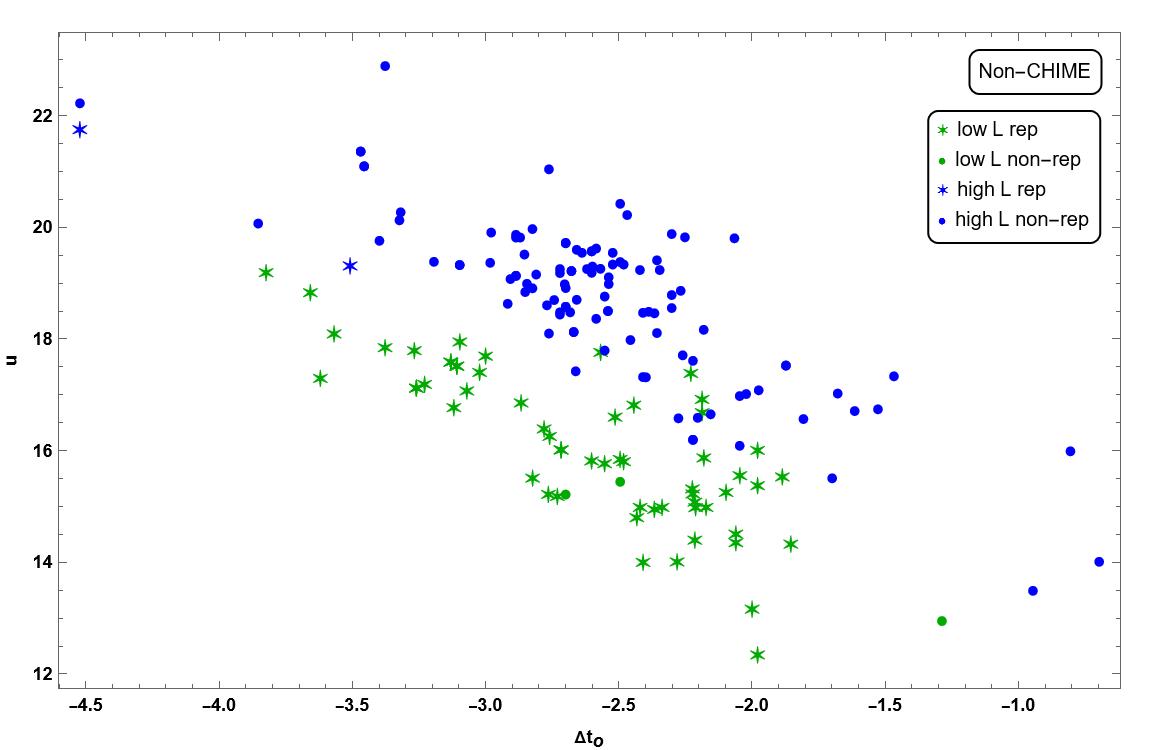}}
    \caption{ (a) A scatter plot of $E$ versus $L_{300}$ shows that they  are statistically correlated with each other. (b) The energy density $u$ against observed duration $\Delta t_O$ shows segregation between the repeaters and the non-repeaters.}
    \label{fig4}
\end{figure}

\begin{figure}[!ht]
    \centering
    \subfigure[]{\includegraphics[width=0.44\textwidth]{ 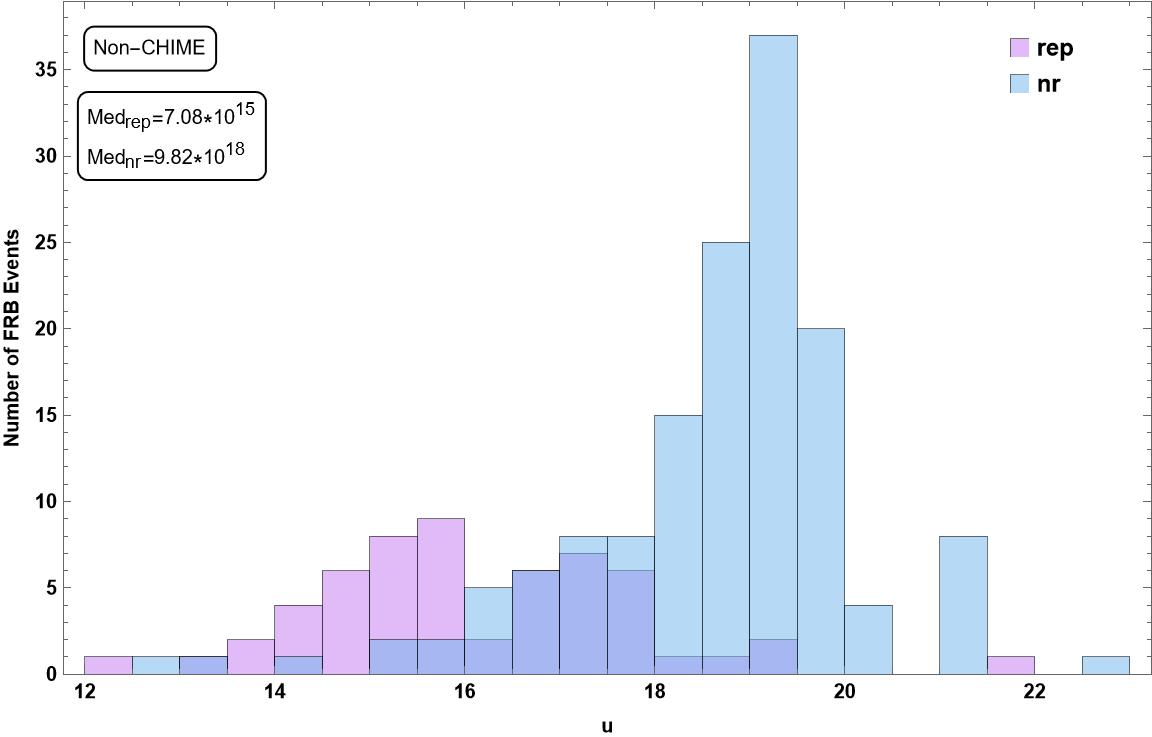}}
    \subfigure[]{\includegraphics[width=0.44\textwidth]{ 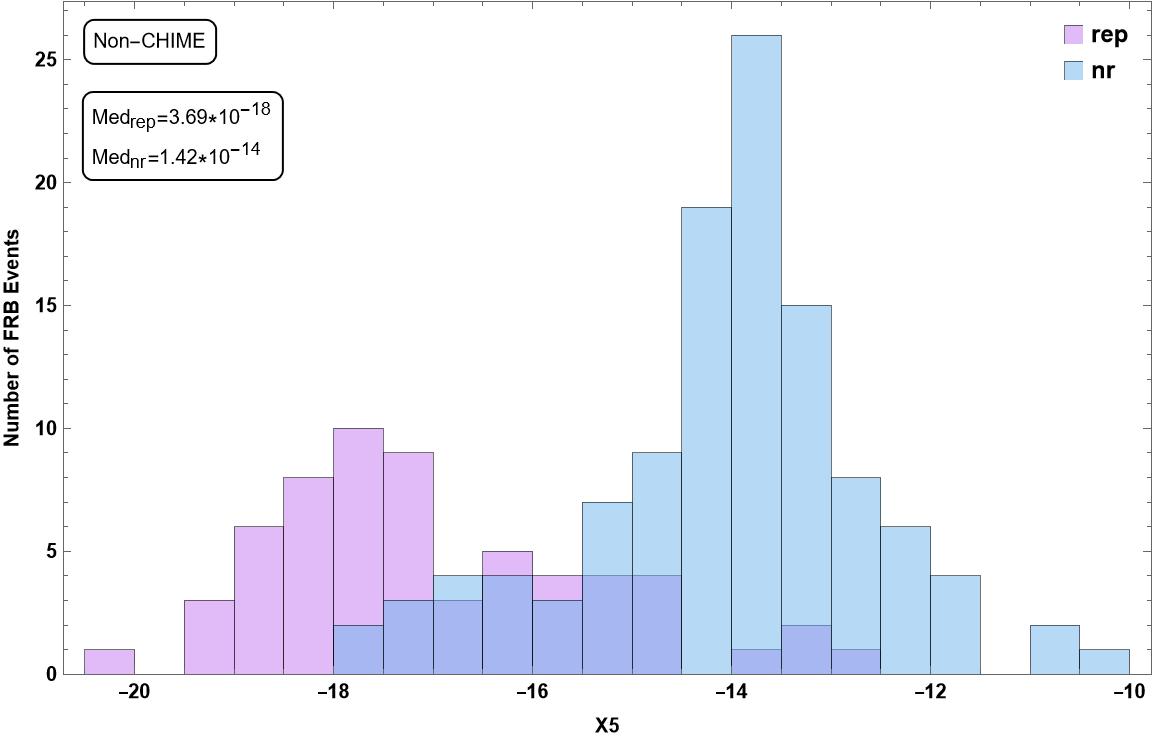}} 
    \caption{ (a) A histogram of the energy density $u$  shows that the repeaters tend to have lower values of $u$ in comparison with that of the one-off FRBs. (b) A histogram of the dimensionless  $X5$ also shows an appreciable segregation of repeaters from the one-off FRBs.}
    \label{fig5}
\end{figure}

% \begin{figure}[!ht]
%     \centering
%     \subfigure[]{\includegraphics[width=0.44\textwidth]{ s7.jpg}}
%     \subfigure[]{\includegraphics[width=0.44\textwidth]{ hX5.jpg}}
%     \caption{(a)  A scatter plot of the dimensionless  $X5$ shows marginal segregation  of repeaters and one-off FRBs. (b) Its histogram shows a bi-modality in $X5$  distribution with repeaters tending to have lower values. }
%     \label{ }
% \end{figure}

%\newpage
\begin{figure}[!ht]
    \centering
    \subfigure[]{\includegraphics[width=0.44\textwidth]{ 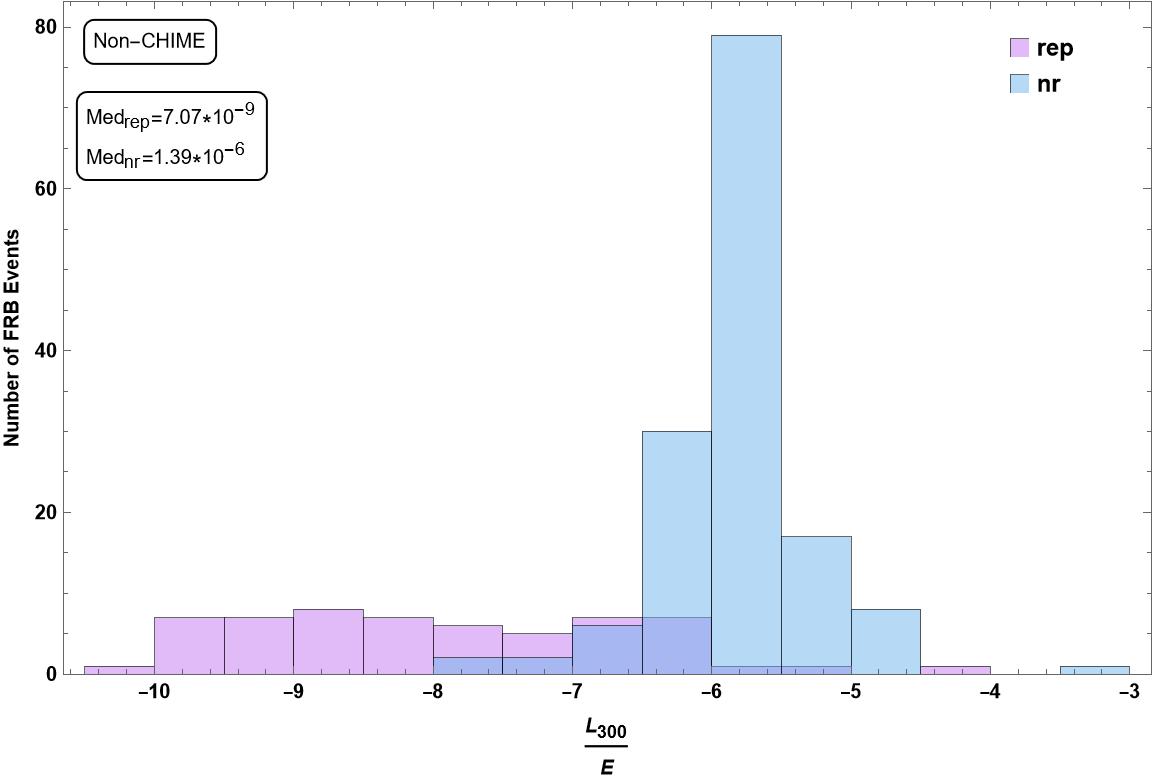}}
    \subfigure[]{\includegraphics[width=0.44\textwidth]{ 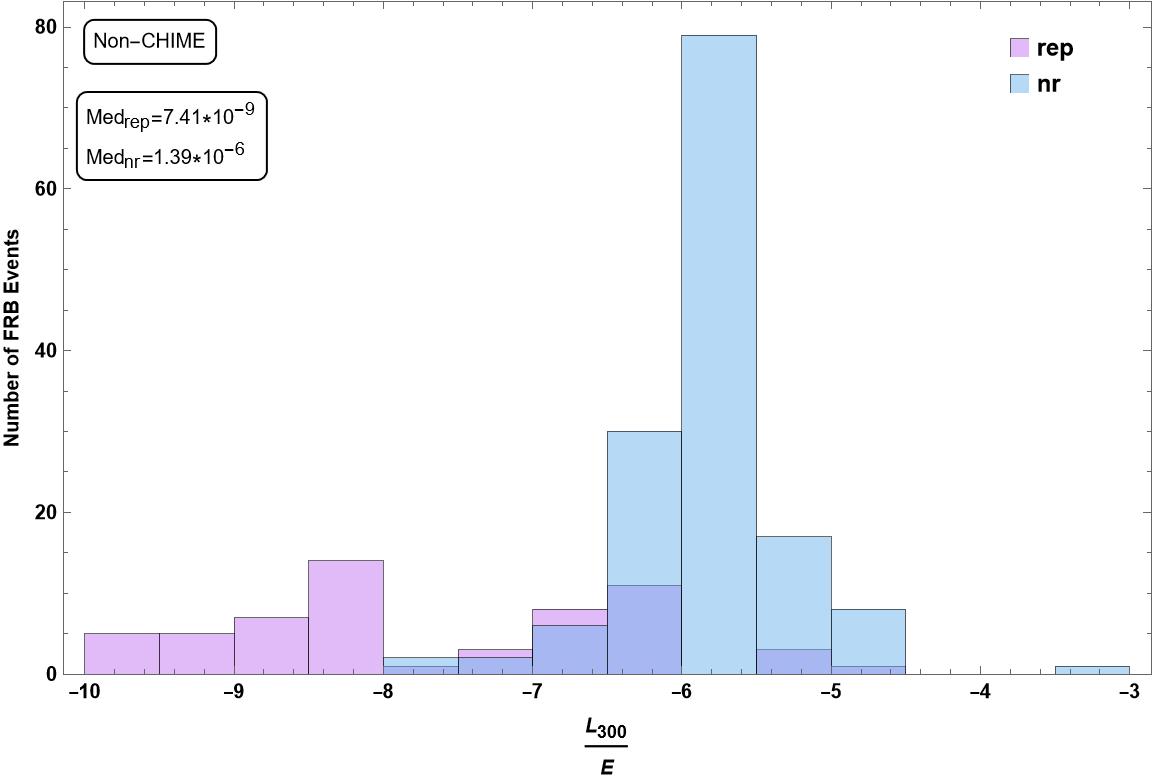}}
    %\vspace{0.1cm}
    \subfigure[]{\includegraphics[width=0.44\textwidth]{ 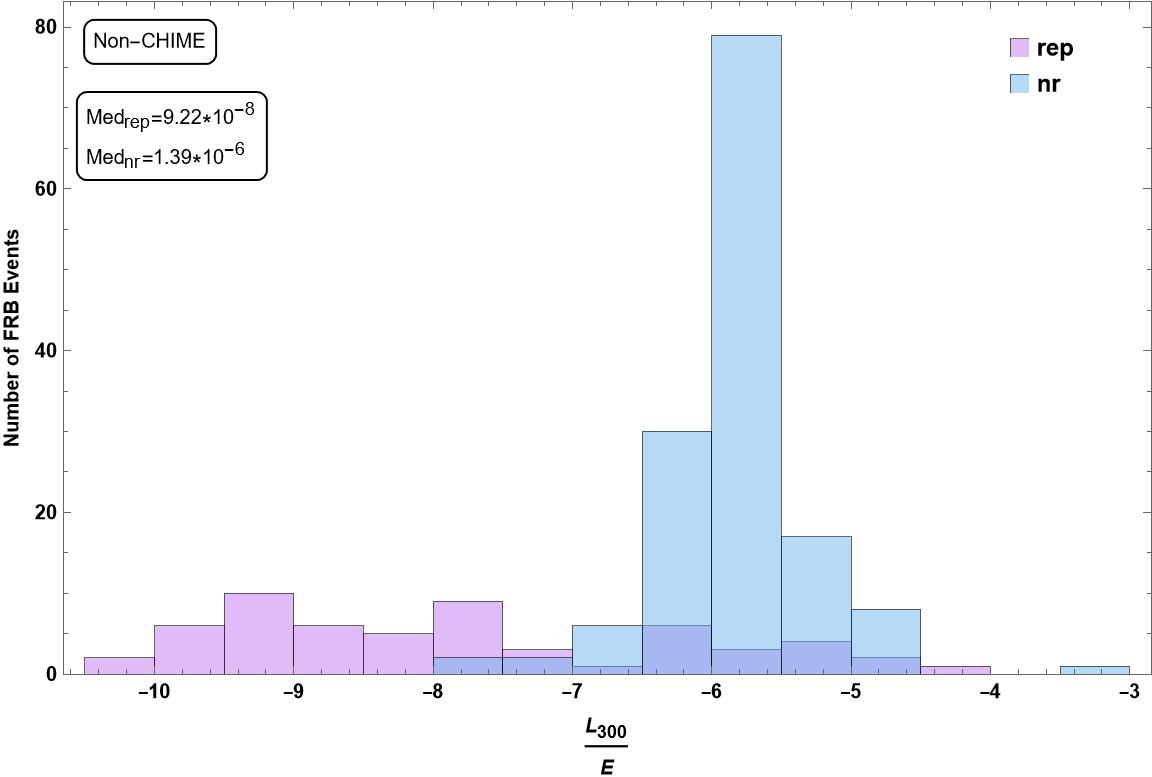}}
    \end{figure}
    \begin{figure}[ht]
    \addtocounter{subfigure}{0}
    \subfigure[]{\includegraphics[width=0.44\textwidth]{ 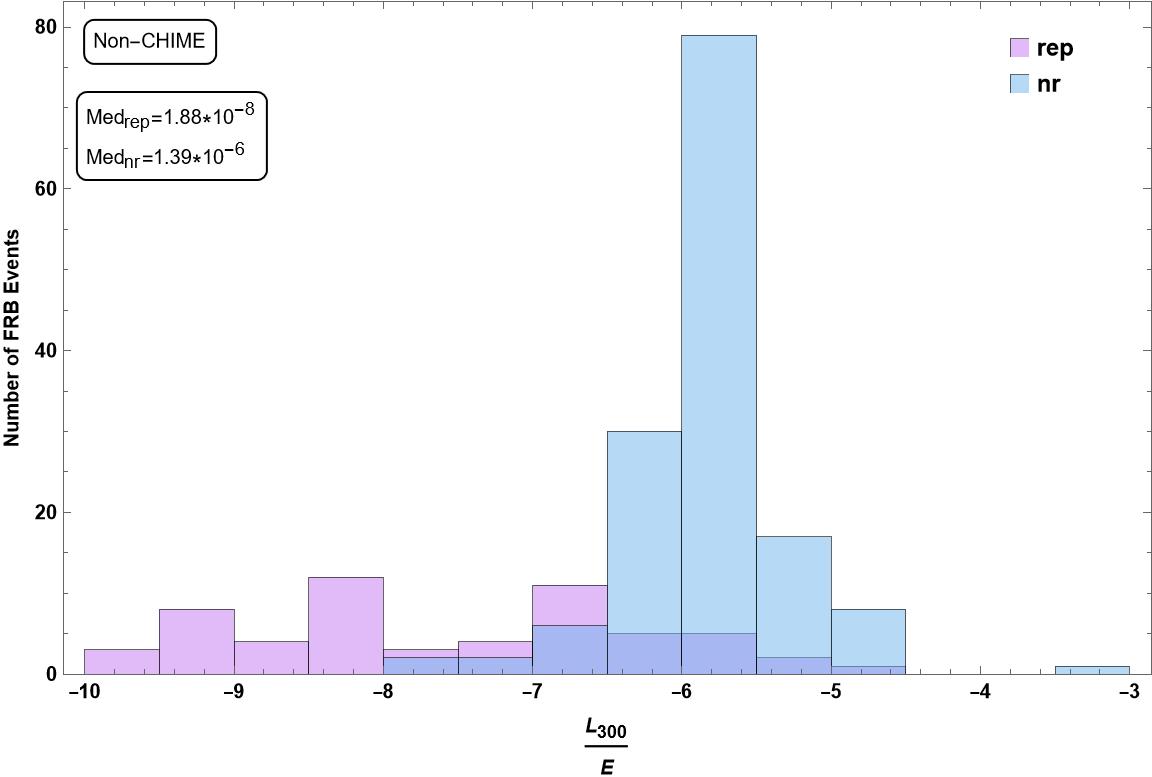}}
    %\vspace{0.1cm}
    \subfigure[]{\includegraphics[width=0.44\textwidth]{ 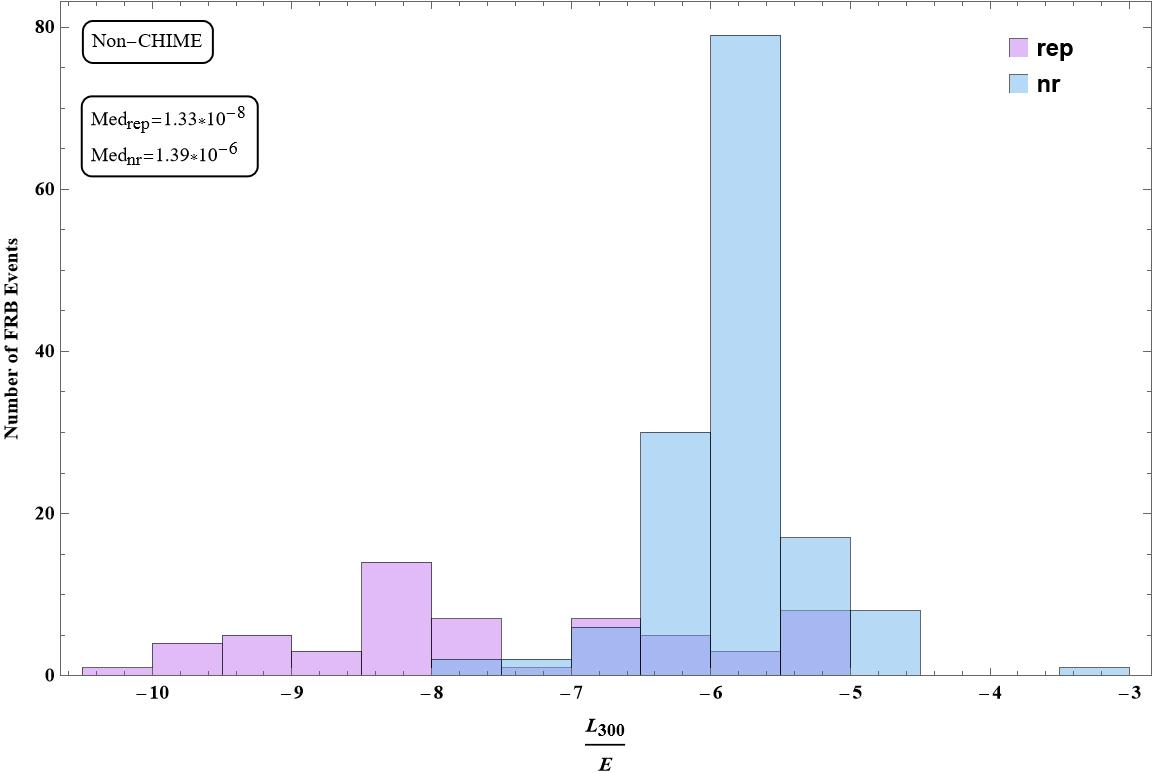}}
    \subfigure[]{\includegraphics[width=0.44\textwidth]{ 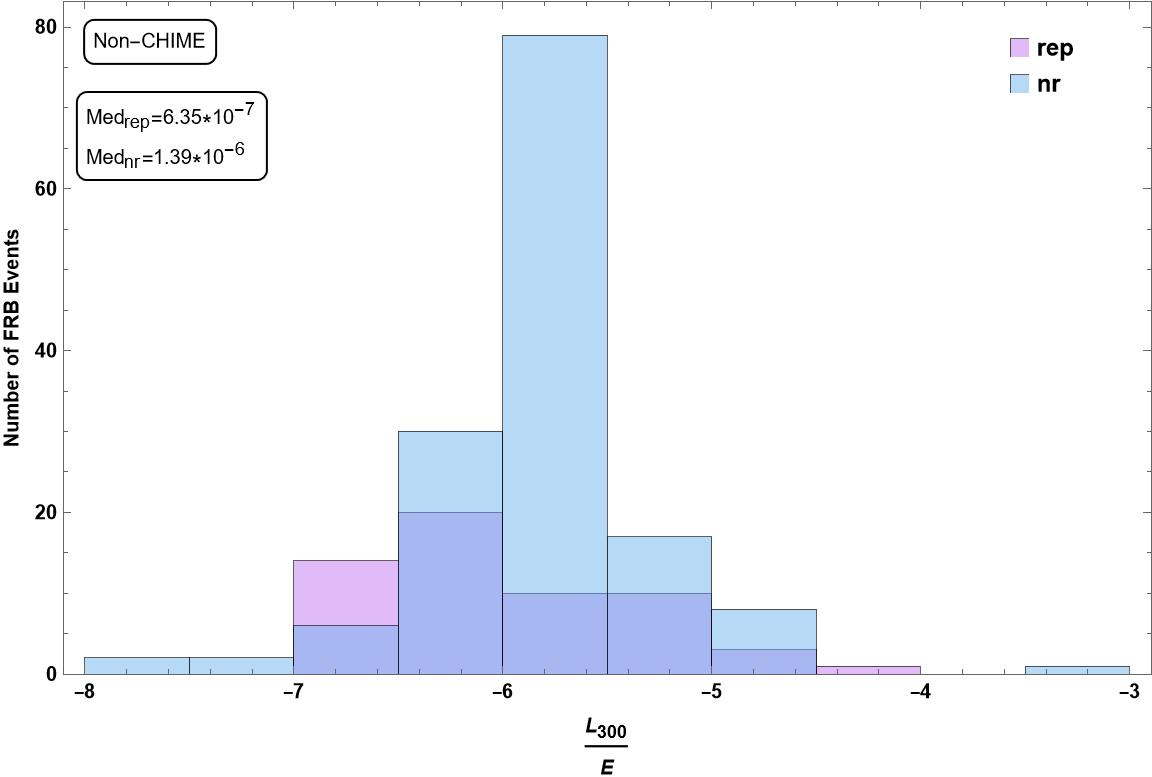}}
    \caption{Histograms of the ratio of Luminosity density at 300 MHz to  Energy for non-CHIME FRBs show 
  bimodal behavior when repeaters are assigned six randomly generated values of the spectral index. The last figure corresponds to $\alpha$=1.5 assigned to all the repeaters.}
    \label{fig6}
\end{figure}
%\vspace{0.1cm}
%

\begin{figure}[!ht]
    \centering
    \subfigure[]{\includegraphics[width=0.44\textwidth]{ 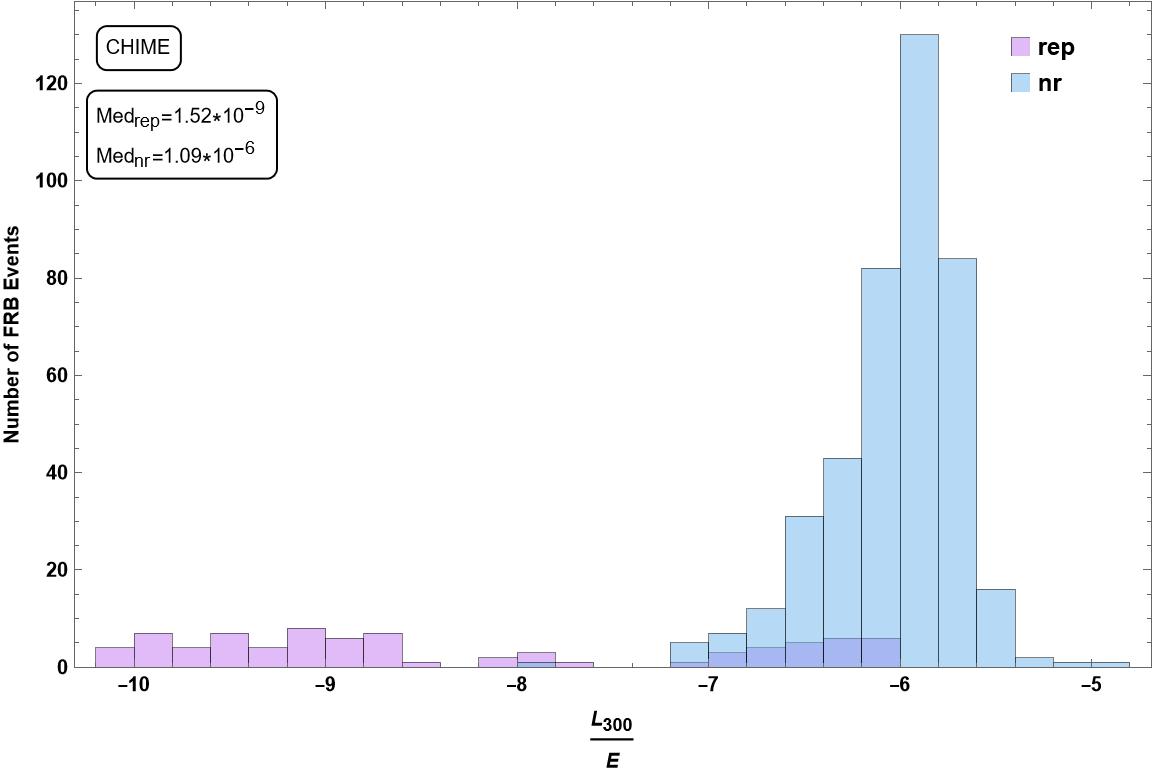}}
    \subfigure[]{\includegraphics[width=0.44\textwidth]{ 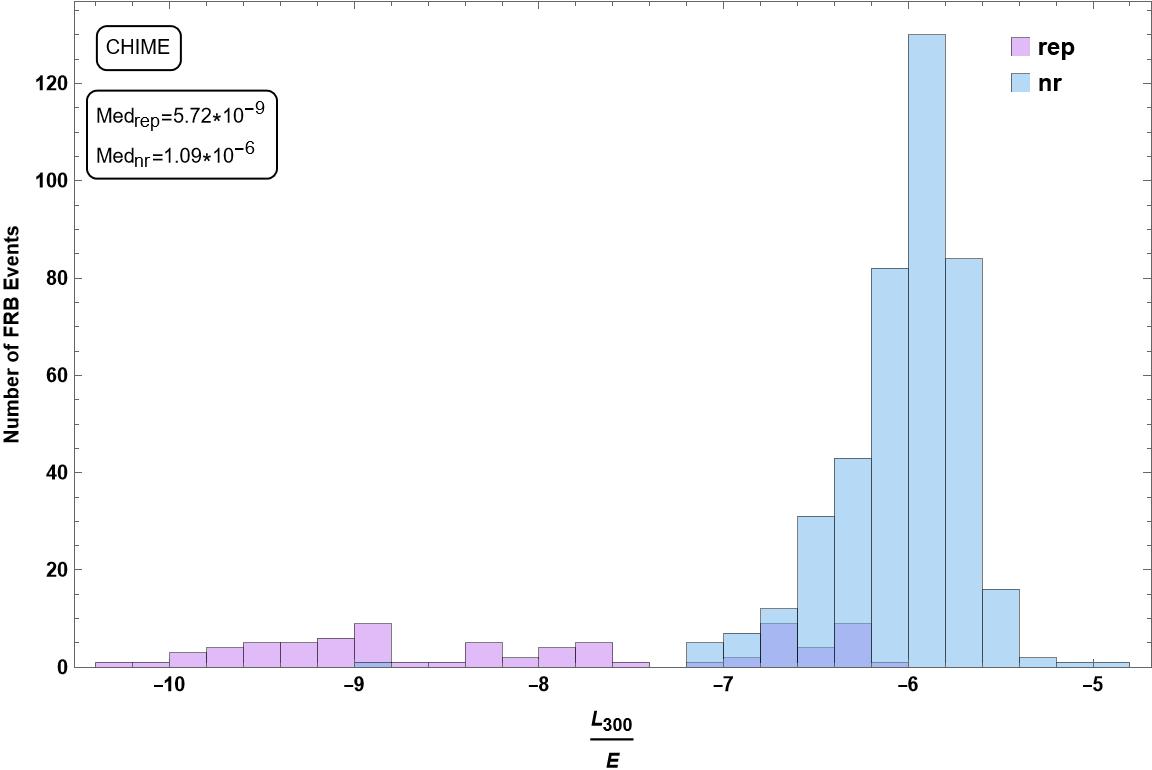}}\\
    %\vspace{0.1cm}
    \subfigure[]{\includegraphics[width=0.44\textwidth]{ 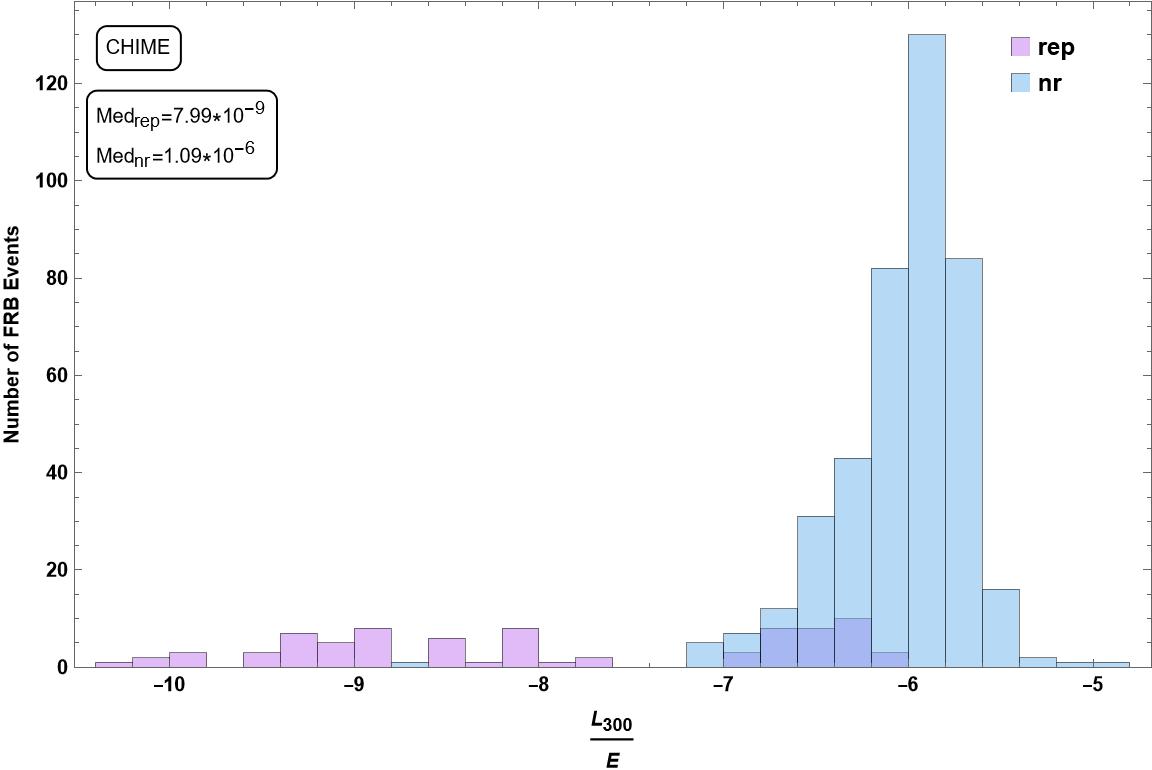}}
    \end{figure}
    \begin{figure}[ht]
    \addtocounter{subfigure}{0}
    \subfigure[]{\includegraphics[width=0.44\textwidth]{ 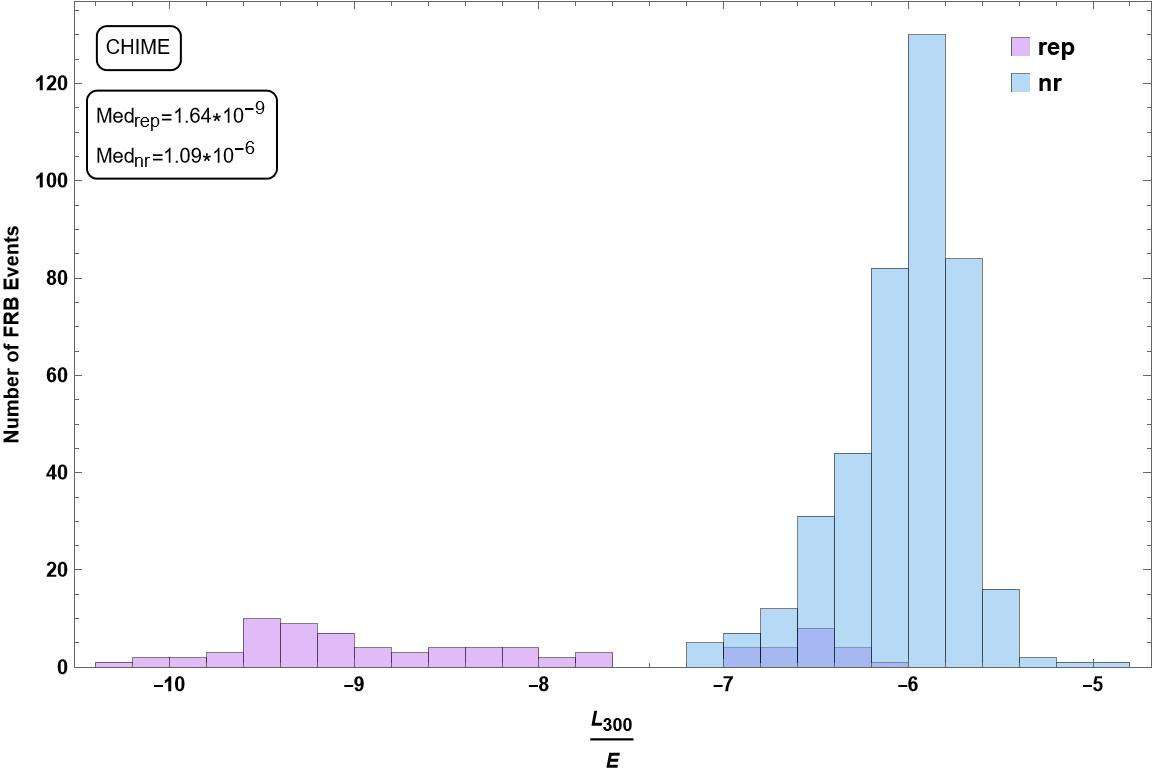}}
    %\vspace{0.1cm}
    \subfigure[]{\includegraphics[width=0.44\textwidth]{ 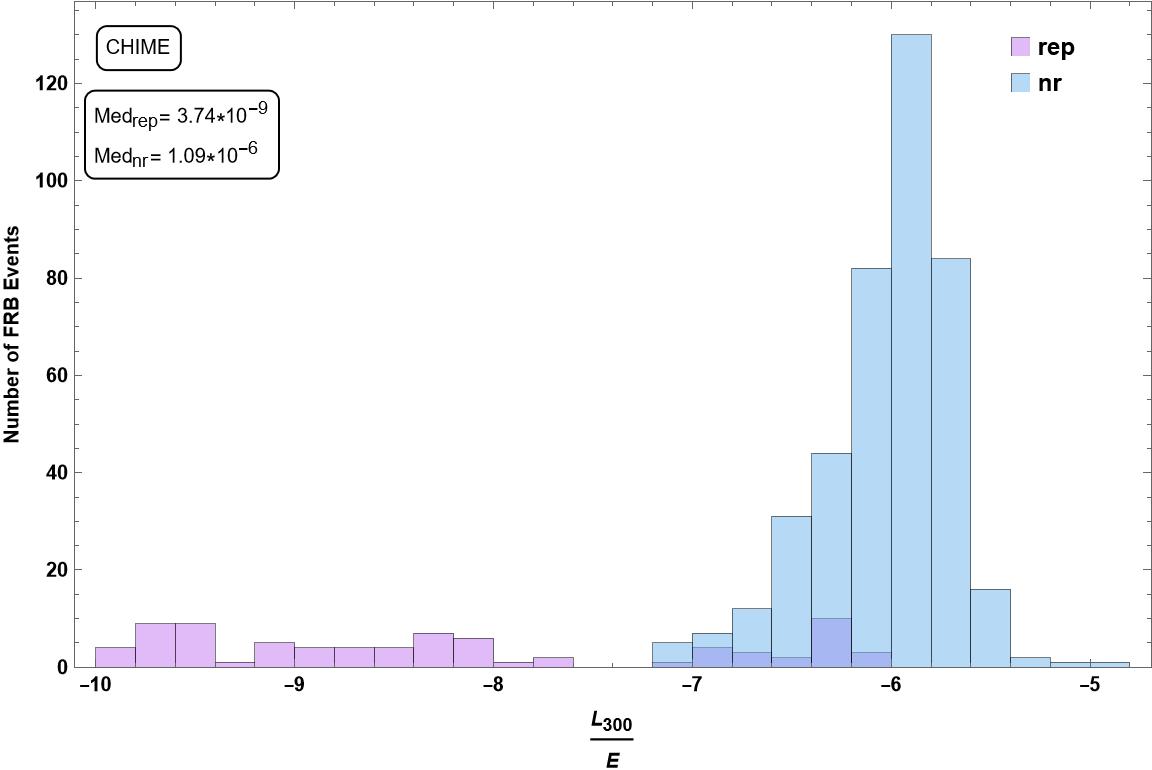}}
    \subfigure[]{\includegraphics[width=0.44\textwidth]{ 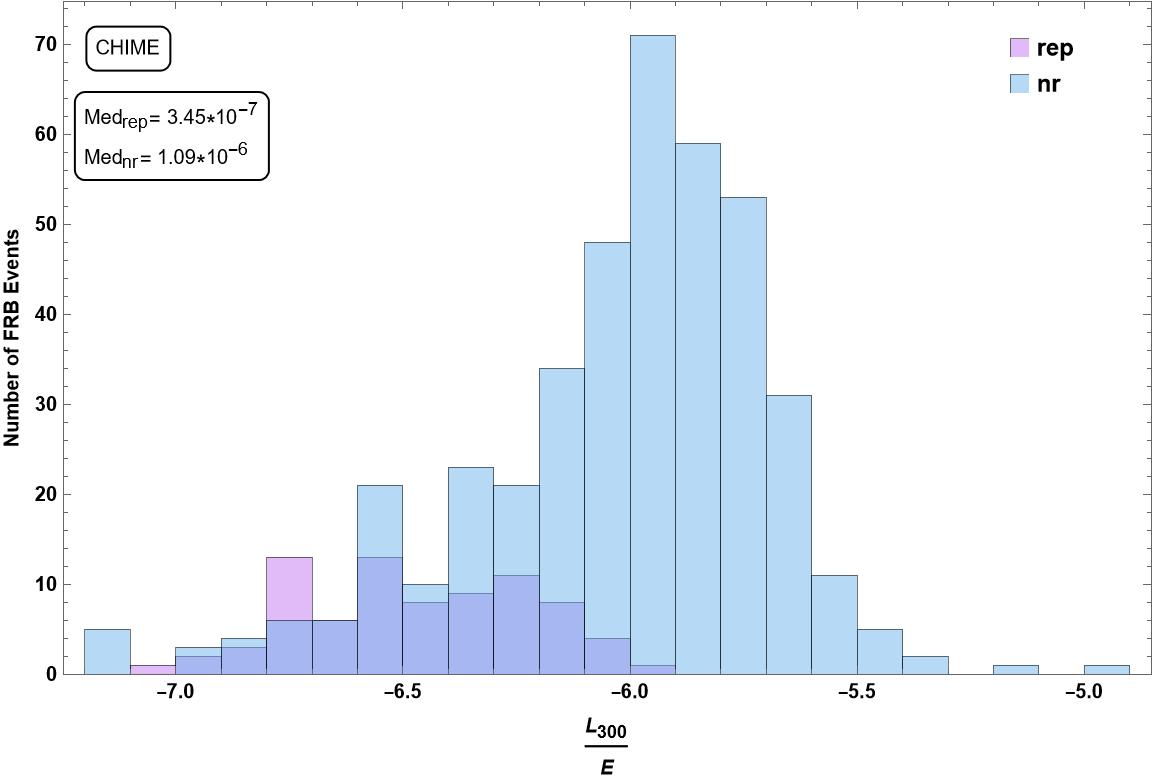}}
    \caption{ For CHIME FRBs, histograms of $L_{300}/E$ show similar bimodality when repeaters are given six randomly generated values of the spectral index. The last histogram corresponds to $\alpha$=1.5 being set for all the repeaters.  }
    \label{fig7}
\end{figure}

\begin{figure}[!ht]
    \centering
    \subfigure[]{\includegraphics[width=0.44\textwidth]{ 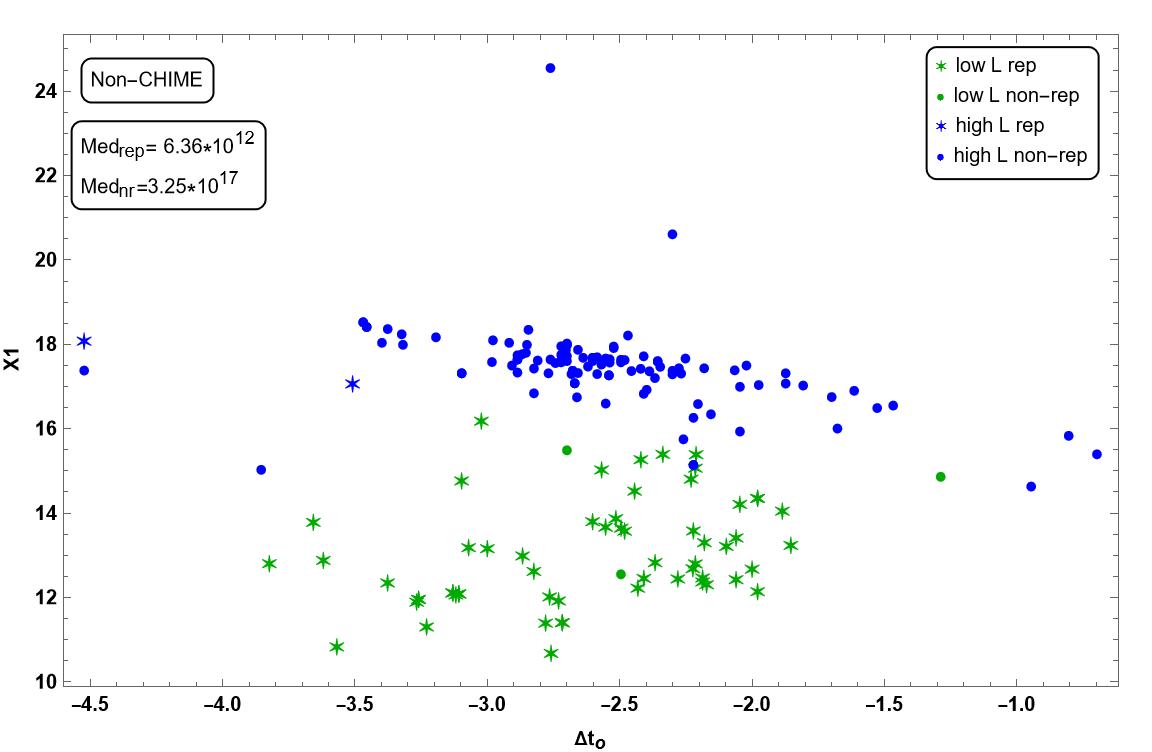}}
    \subfigure[]{\includegraphics[width=0.44\textwidth]{ 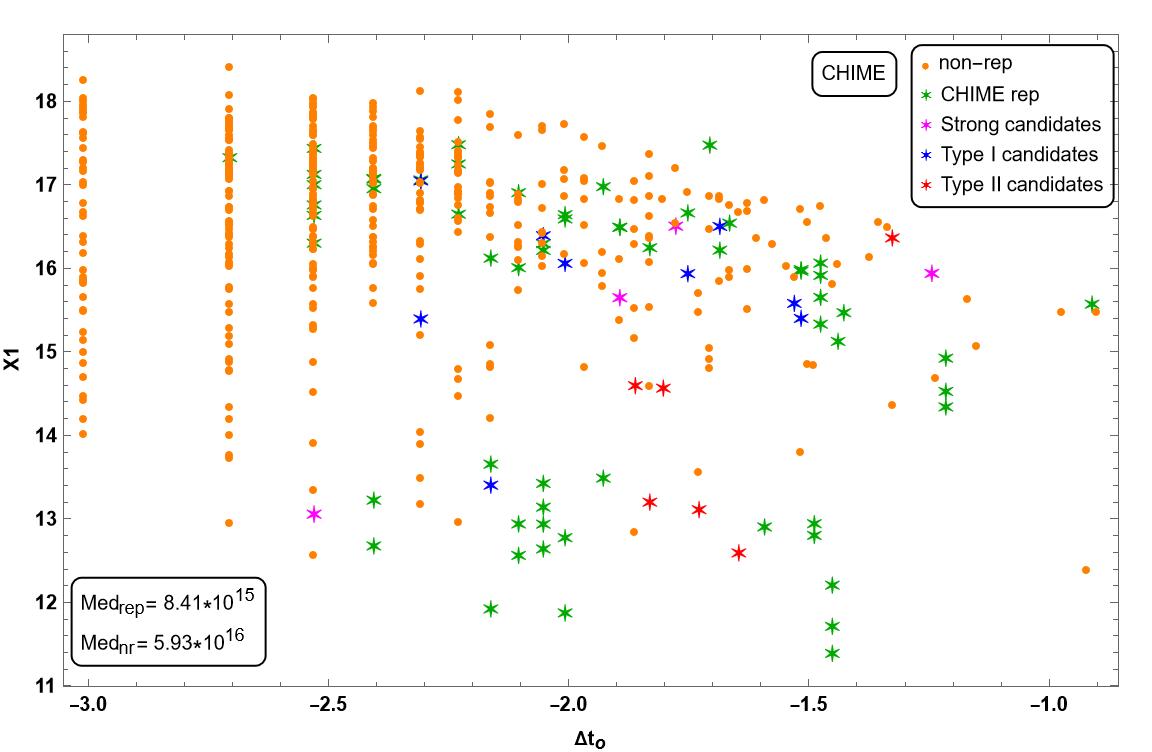}} 
    \caption{ Scatter diagrams of $X1$ against the observed FRB duration for (a) non-CHIME and (b) CHIME. For both, repeaters tend to have lower values of $X1$.}
    \label{fig8}
\end{figure}

% \begin{figure}[!ht]
%     \centering
%     \subfigure[]{\includegraphics[width=0.44\textwidth]{ s2.jpg}}
%     \subfigure[]{\includegraphics[width=0.44\textwidth]{ s2.jpg}} 
%     \caption{(a) blah (b) blah}
%     \label{fig:foobar}
% \end{figure}
\begin{figure}[!ht]
    \centering
    \subfigure[]{\includegraphics[width=0.44\textwidth]{ 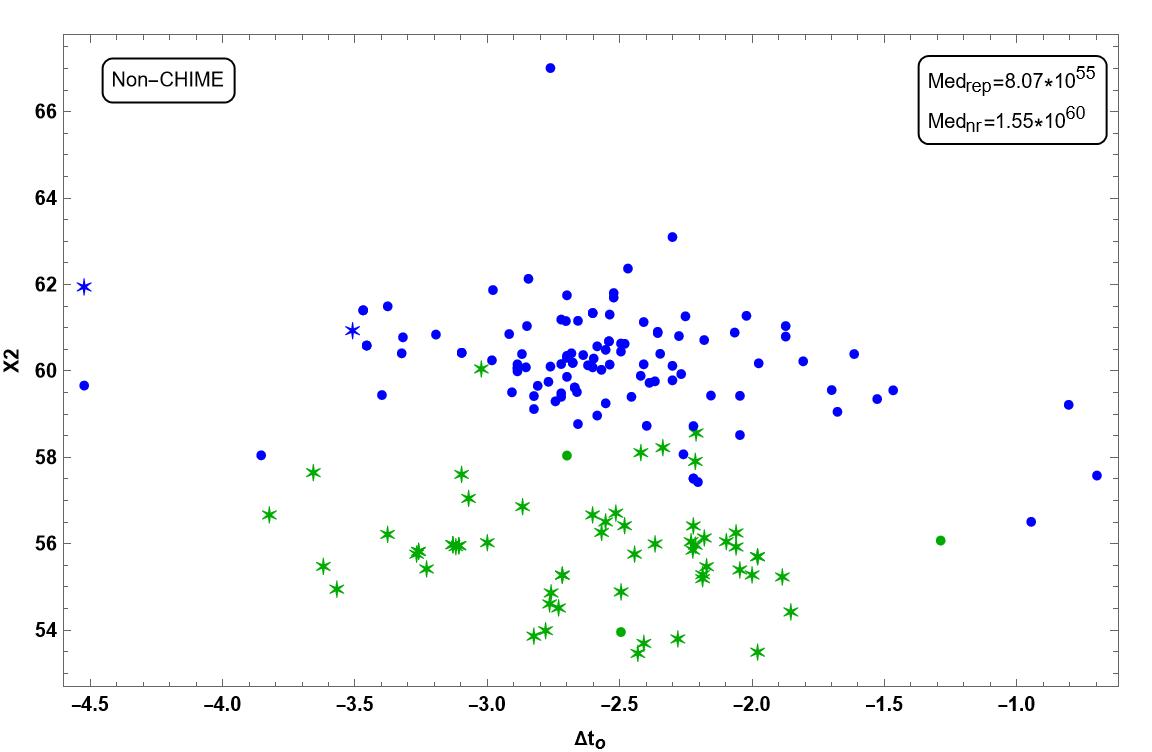}}
    \end{figure}
    \begin{figure}[ht]
    \addtocounter{subfigure}{0}
    \subfigure[]{\includegraphics[width=0.44\textwidth]{ 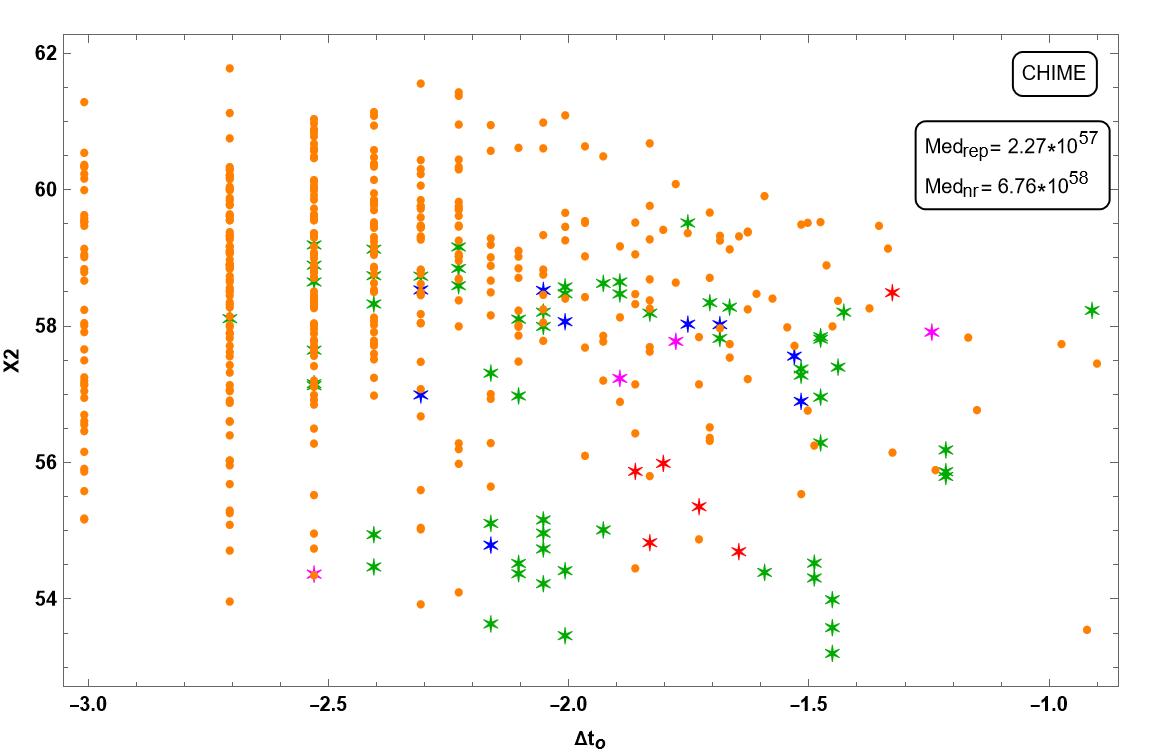}} 
    \caption{ Scatter plots of $X2$ versus $\Delta t_O$ for (a) non-CHIME and (b) CHIME both show similar behavior with repeaters tending to have marginally smaller values of $X2$. }
    \label{fig9}
\end{figure}

% \begin{figure}[!ht]
%     \centering
%     \subfigure[]{\includegraphics[width=0.44\textwidth]{ s4.jpg}}
%     \subfigure[]{\includegraphics[width=0.44\textwidth]{ s4.jpg}} 
%     \caption{The plots d(a)  (b)  The indicators are provided in Figure \ref{ncEL} and \ref{s1}  for both CHIME and non-CHIME.}
%     \label{fig:foobar}
% \end{figure}
\begin{figure}[!ht]
    \centering
    \subfigure[]{\includegraphics[width=0.44\textwidth]{ 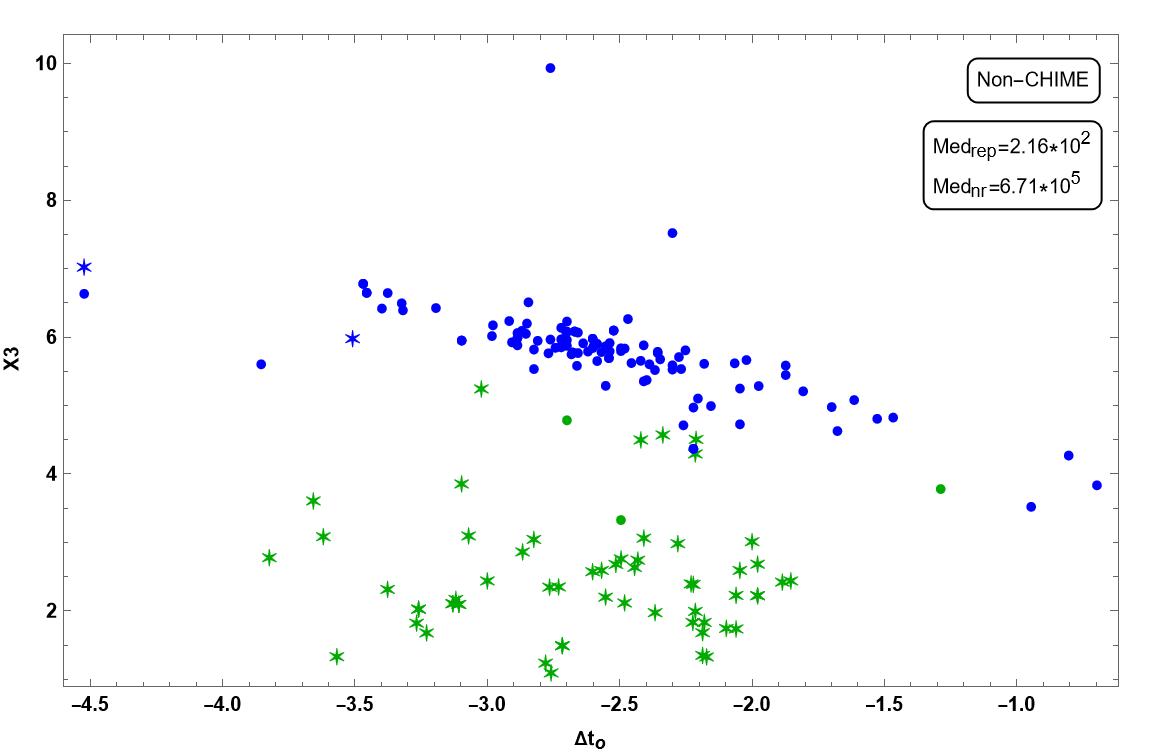}}
    \subfigure[]{\includegraphics[width=0.44\textwidth]{ 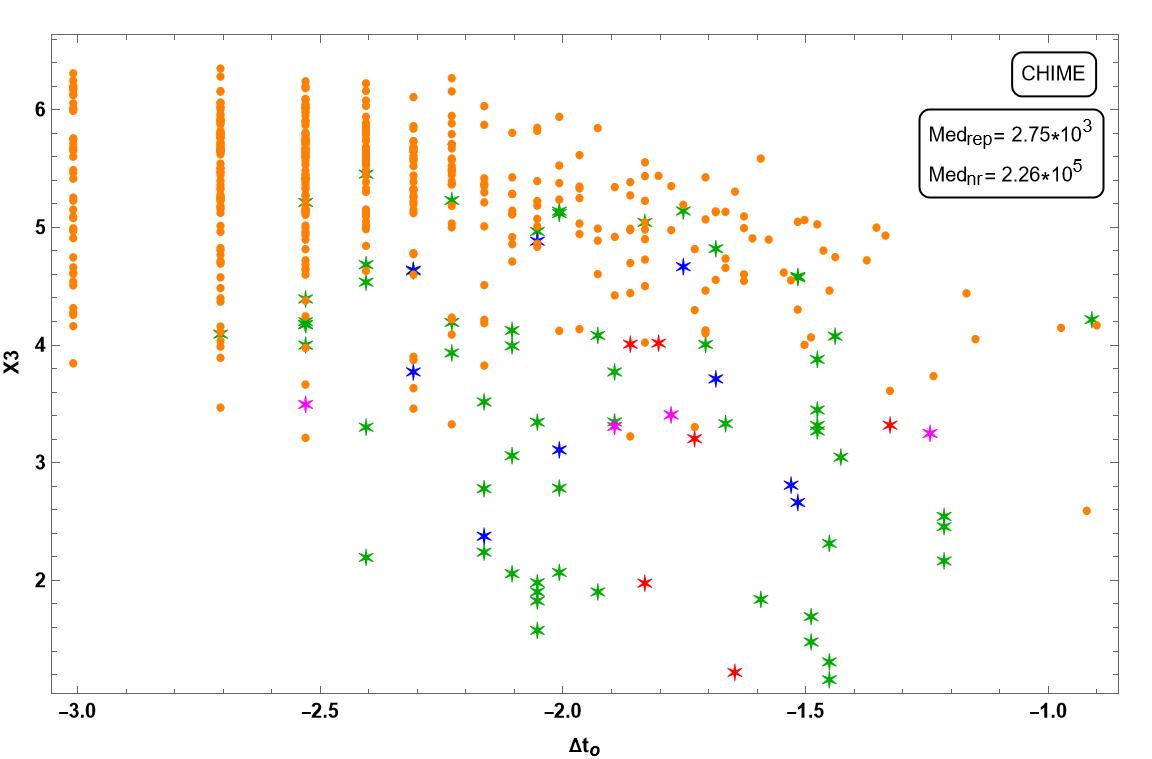}} 
    \caption{ Scatter diagrams of $X3$ against $\Delta t_O$ for (a) non-CHIME and (b) CHIME both show similar trends, in that, repeaters tend to have lower values of $X3$.}
    \label{fig10}
\end{figure}

\begin{figure}[!ht]
    \centering
    \subfigure[]{\includegraphics[width=0.44\textwidth]{ 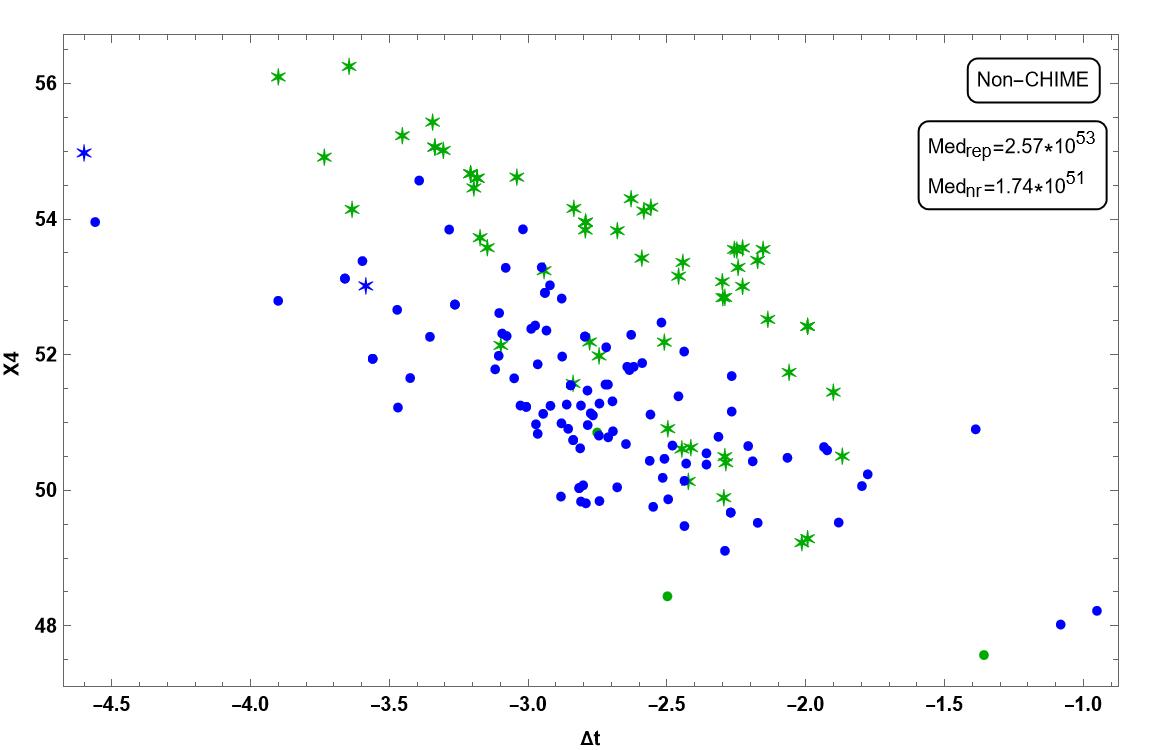}}
    \subfigure[]{\includegraphics[width=0.44\textwidth]{ 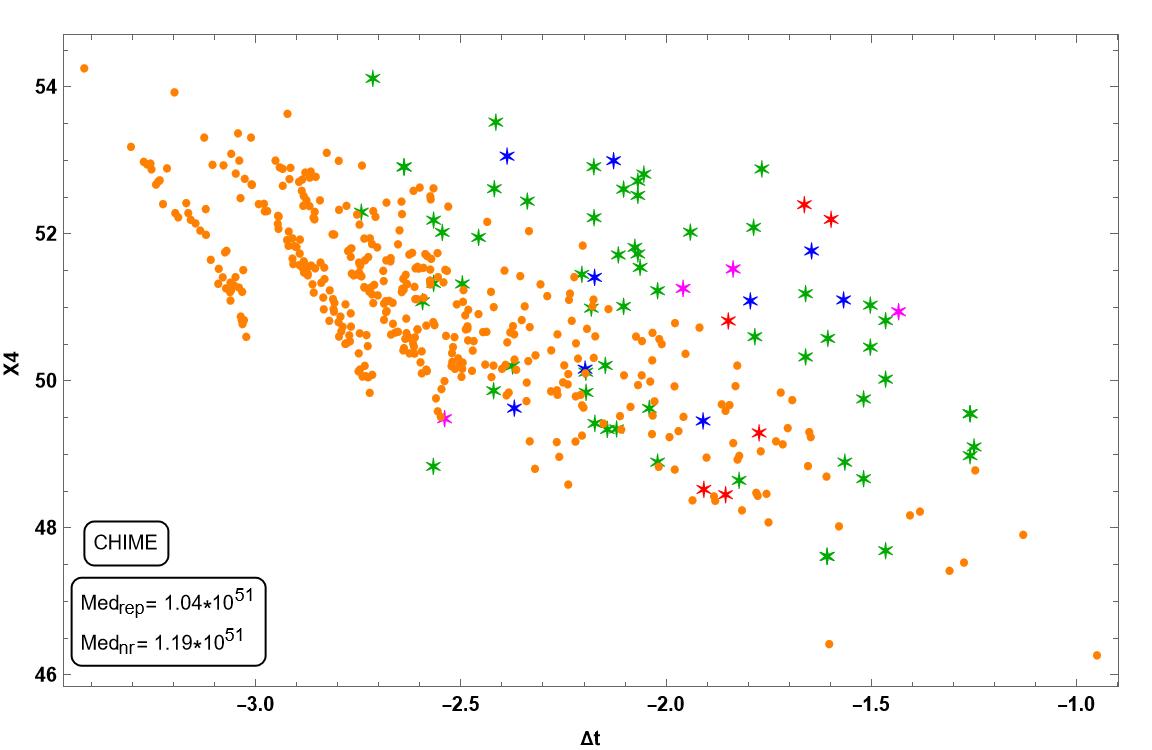}} 
    \caption{ The dimensionless quantity $X4$ when plotted versus intrinsic duration $\Delta t$  show for (a) non-CHIME as well as (b) CHIME that $X4$ for the repeaters tend to lie to the right of those for the non-repeaters. }
    \label{fig11}
\end{figure}

\begin{figure}[!ht]
    \centering
    \subfigure[]{\includegraphics[width=0.44\textwidth]{ 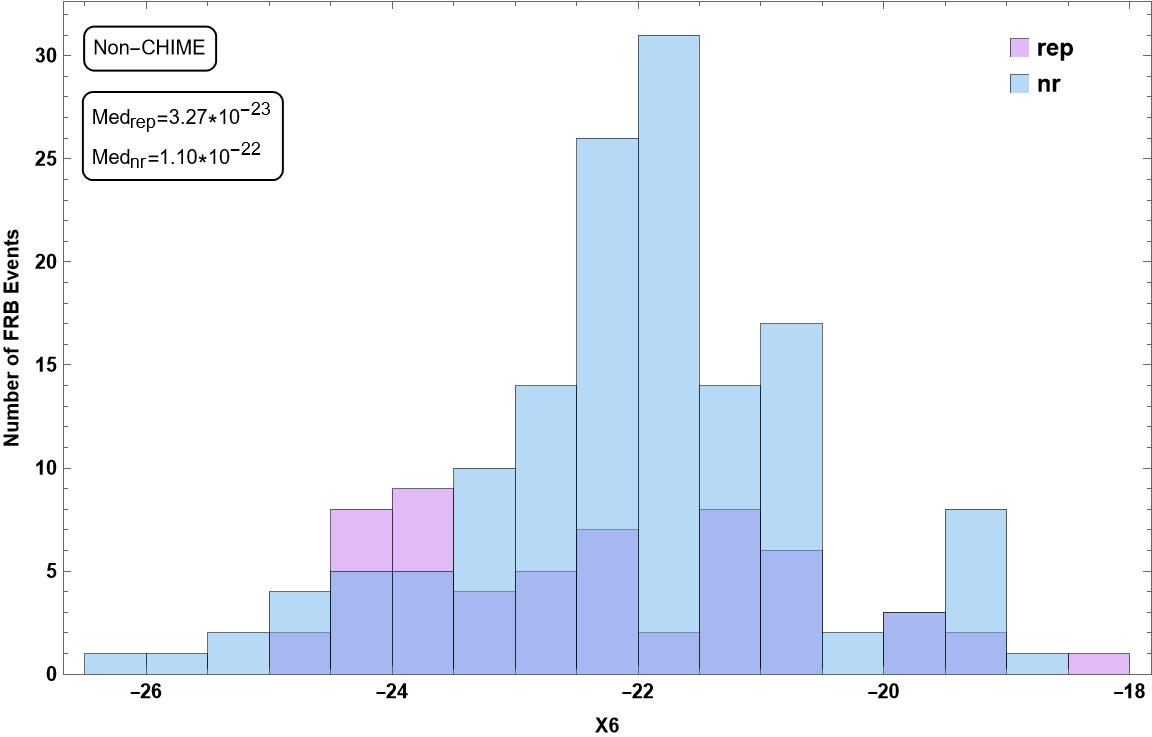}}
    \end{figure}
    \begin{figure}[ht]
    \addtocounter{subfigure}{0}
    \subfigure[]{\includegraphics[width=0.44\textwidth]{ 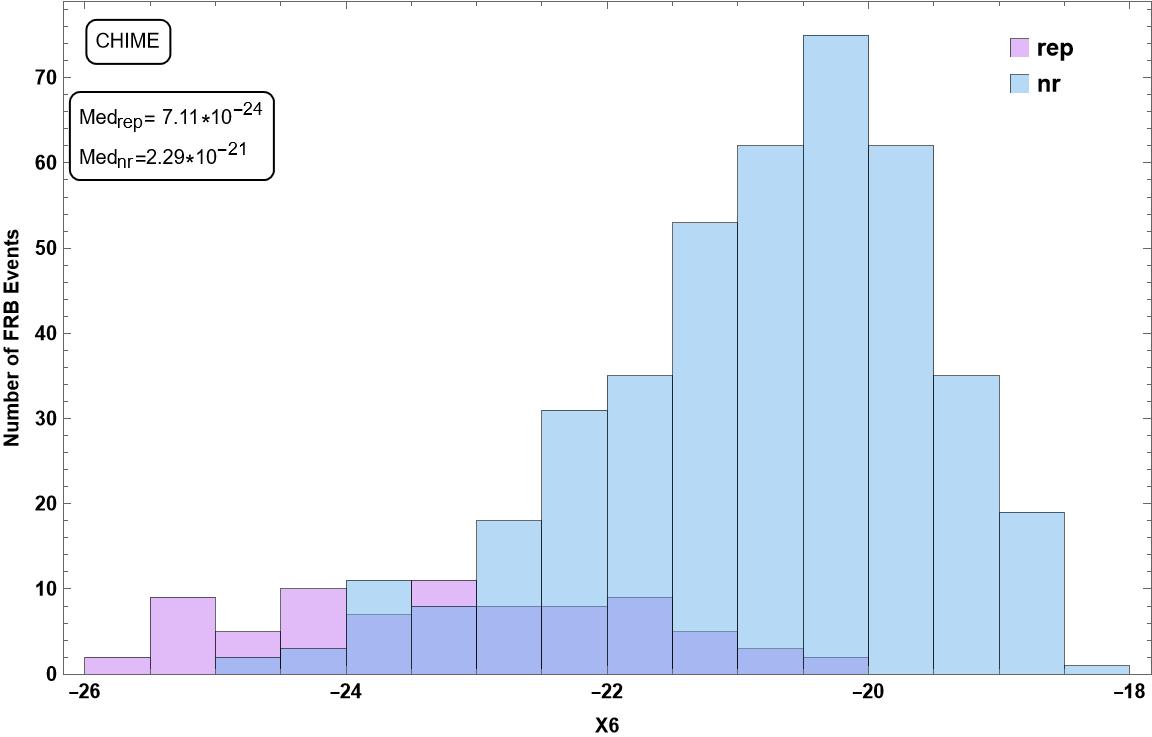}} 
    \caption{The CHIME histogram of the ratio $X6 \equiv k_B\ T_b/E$ in (b)  shows considerable separation in the peaks for REPs and NRs. In the case of (a) non-CHIME FRBs, the difference is marginal.} 
    \label{fig12}
\end{figure}

\begin{figure}[!ht]
    \centering
    \subfigure[]{\includegraphics[width=0.44\textwidth]{ 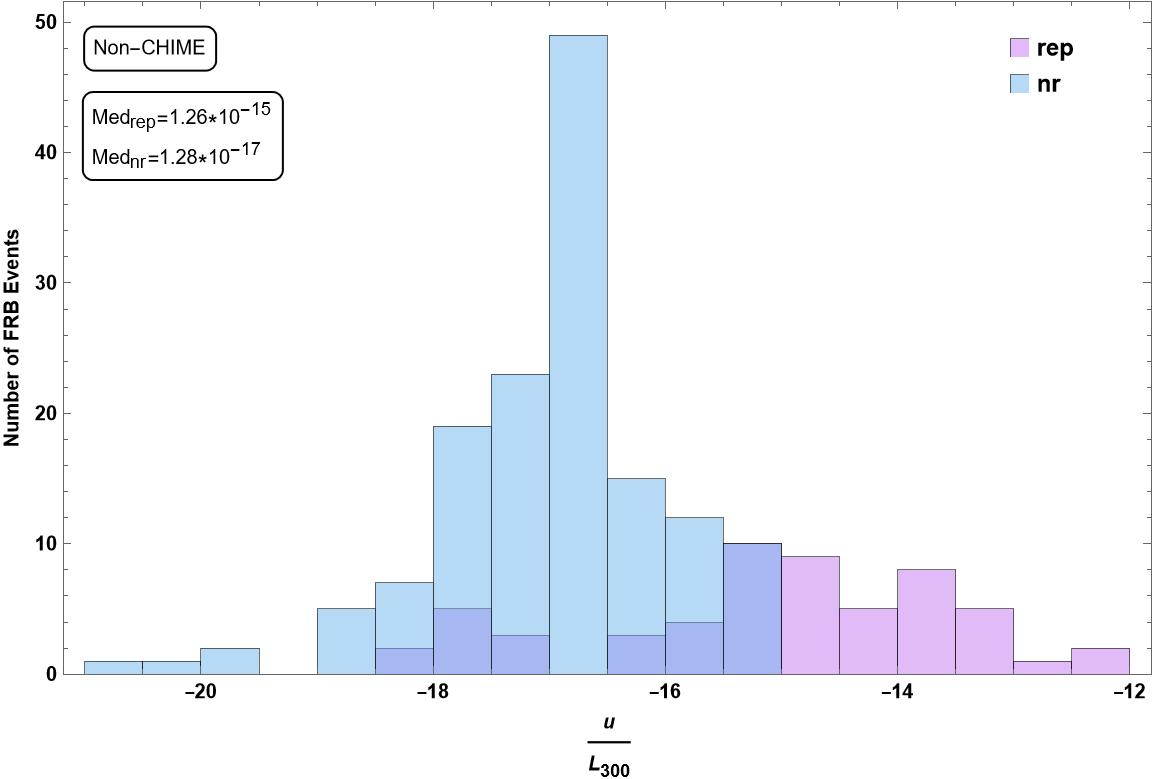}}
    \subfigure[]{\includegraphics[width=0.44\textwidth]{ 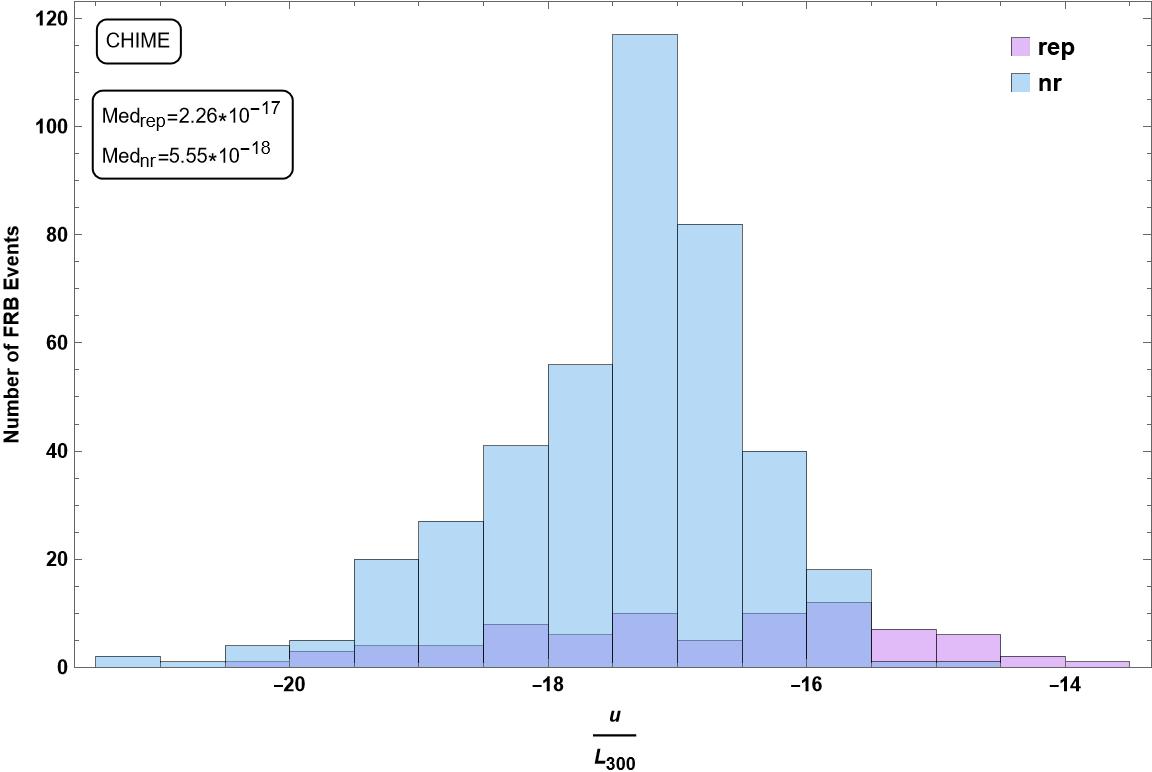}} 
    \caption{ (a) A histogram of the energy density $u/L_{300}$  for non-CHIME shows that the repeaters tend to have higher values in comparison with that of the one-off FRBs. (b) The distribution of $u/L_{300}$ for the CHIME FRBs  shows  marginal segregation of repeaters and one-off FRBs, with repeaters tending to have larger values.}
    \label{fig13}
\end{figure}

\begin{figure}[!ht]
    \centering
    \subfigure[]{\includegraphics[width=0.44\textwidth]{ 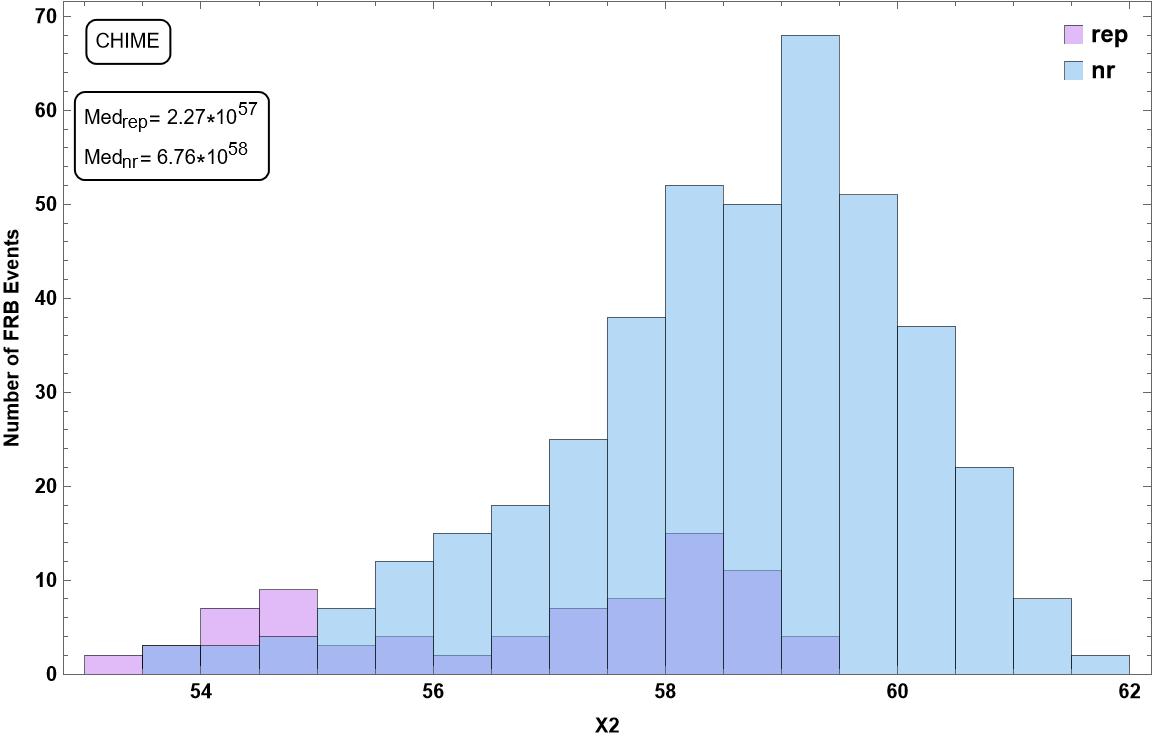}}
    \subfigure[]{\includegraphics[width=0.44\textwidth]{ 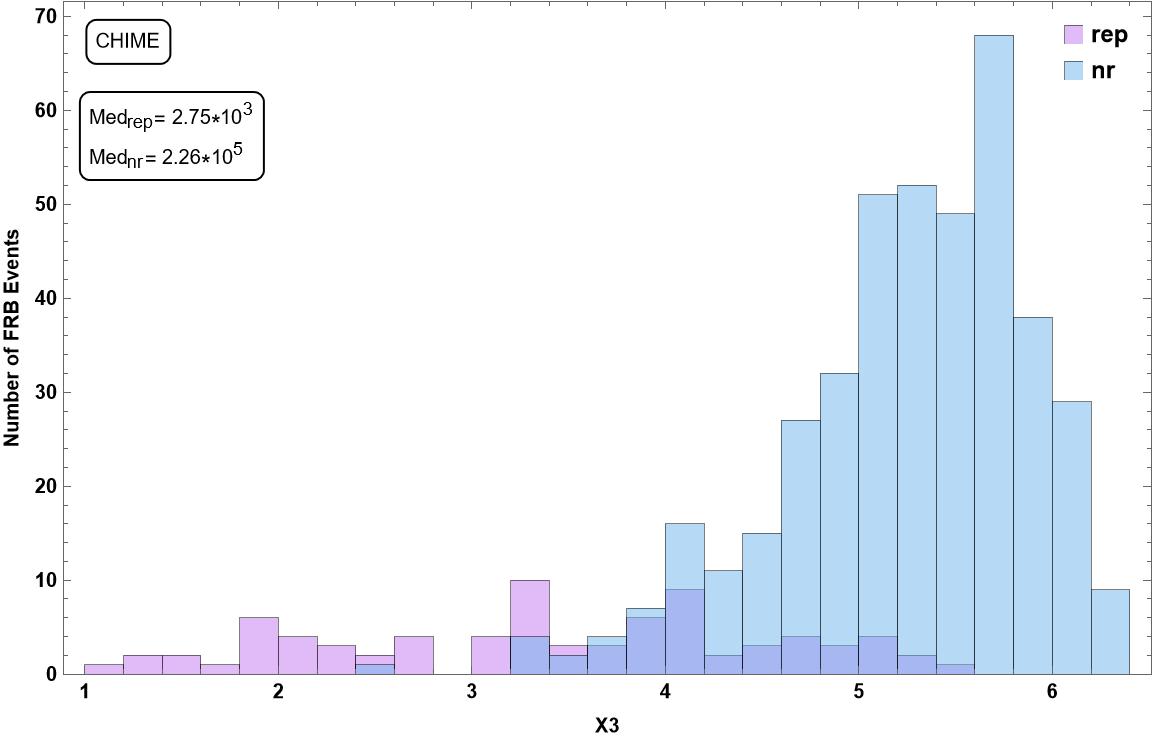}}
    \subfigure[]{\includegraphics[width=0.44\textwidth]{ 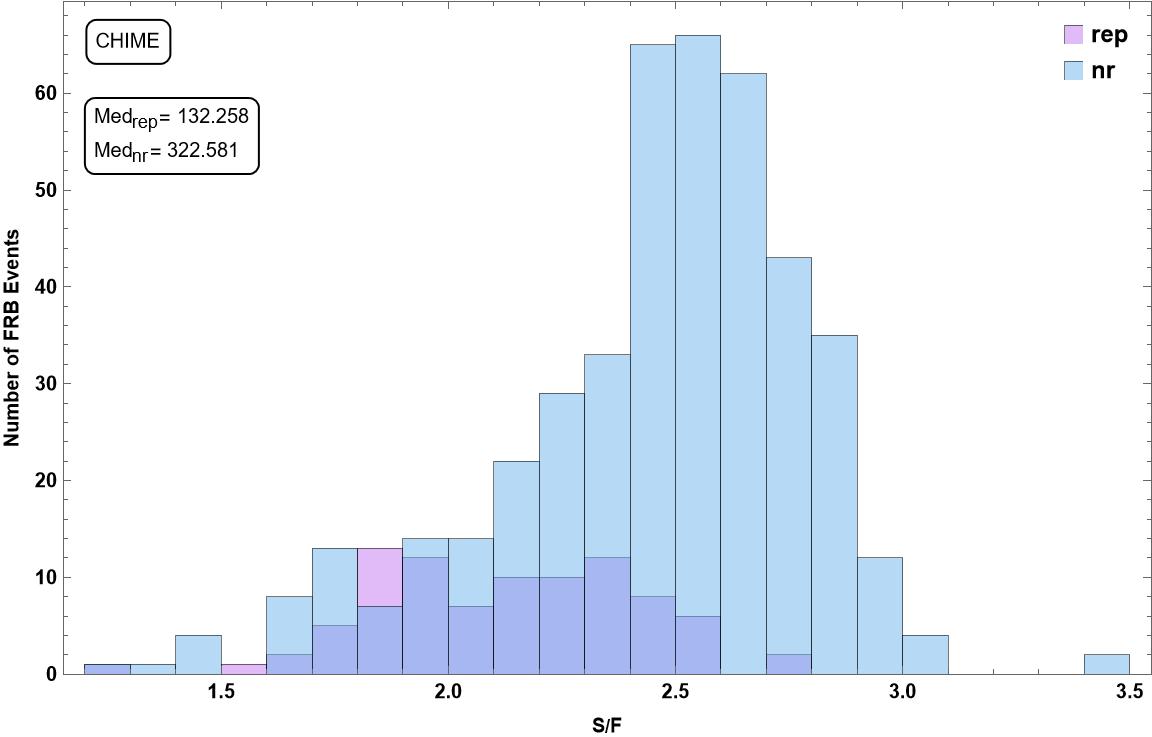}} 
    \caption{ The histograms for (a) $X2 $, (b) $ X3$, and (c) ratio of flux density fluence for CHIME FRBs show segregation in the distributions for repeaters and non-repeaters. } 
    \label{fig14}
\end{figure}

% \begin{figure}[!ht]
%     \centering
%     \subfigure[]{\includegraphics[width=0.44\textwidth]{ ncsX7dm.jpg}}
%     \subfigure[]{\includegraphics[width=0.44\textwidth]{ chsX7dm.jpg}} 
%     \caption{ (a) .}
%     \label{fig.14}
% \end{figure}

% \begin{figure}[!ht]
%     \centering
%     \subfigure[]{\includegraphics[width=0.44\textwidth]{ ncsX7X4.jpg}}
%     \subfigure[]{\includegraphics[width=0.44\textwidth]{ chsX7X4.jpg}} 
%     \caption{ (a) }
%     \label{fig.15}
% \end{figure}

% \begin{figure}[!ht]
%     \centering
%     \subfigure[]{\includegraphics[width=0.44\textwidth]{ ncsX7X8.jpg}}
%     \subfigure[]{\includegraphics[width=0.44\textwidth]{ chsX7X8.jpg}} 
%     \caption{ (a) }
%     \label{fig.16}
% \end{figure}

\begin{figure}[!ht]
    \centering
    % \subfigure[]{\includegraphics[width=0.44\textwidth]{ nchrX7.jpg}}
    % \subfigure[]{\includegraphics[width=0.44\textwidth]{ nchX7.jpg}} 
    \subfigure[]{\includegraphics[width=0.44\textwidth]{ 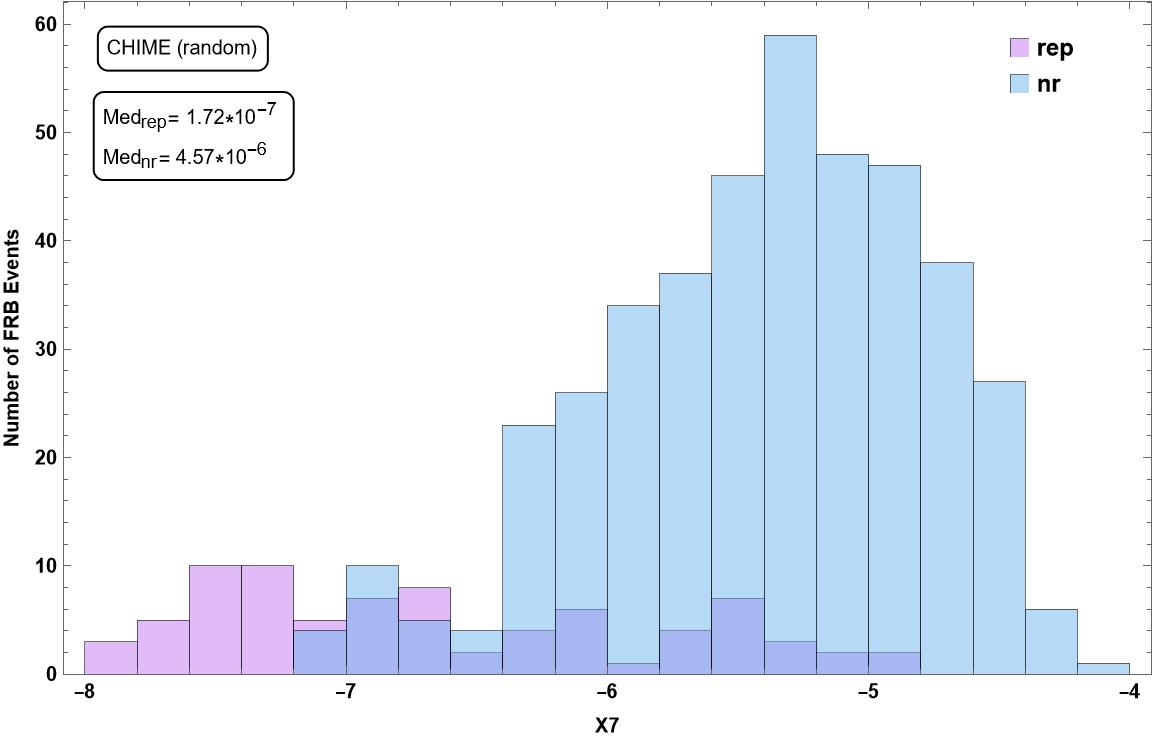}}
    \subfigure[]{\includegraphics[width=0.44\textwidth]{ 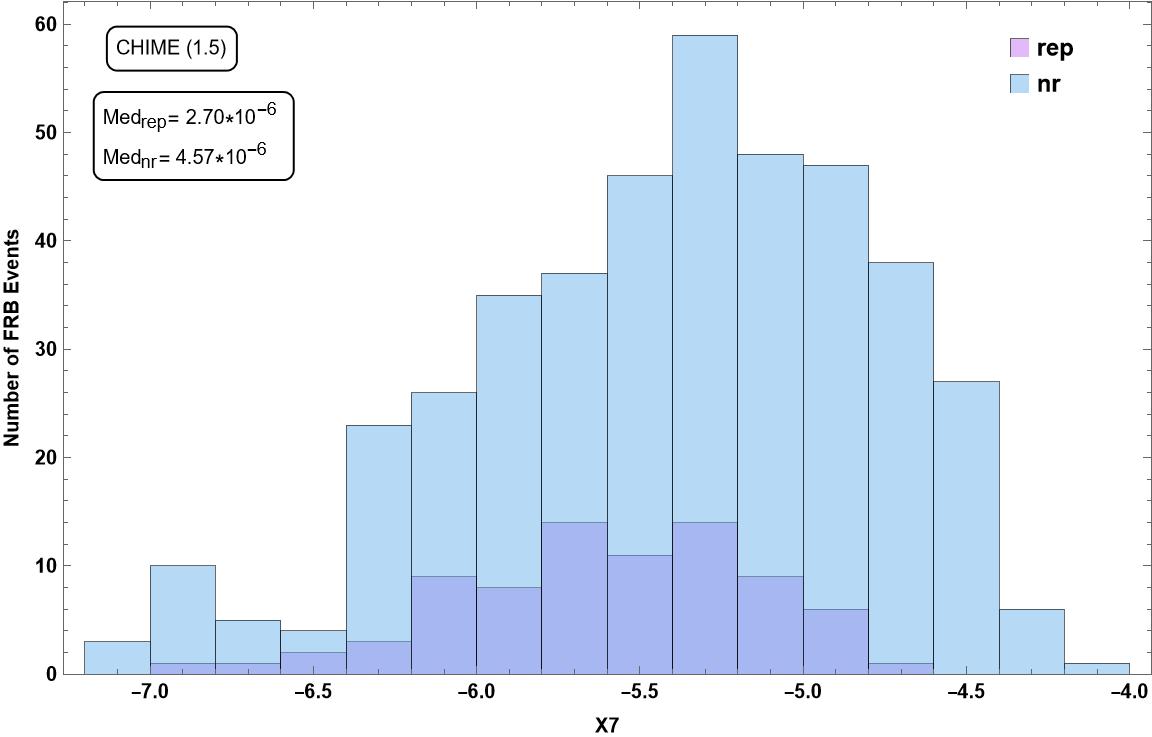}} 
    \caption{Histograms of $X7$ for CHIME FRBs: (a) For a case when random values of $\alpha $ were assigned to the repeaters and (b) when $\alpha = 1.5$ is set for all the repeaters. The median values reflect the difference exhibited by repeaters and non-repeaters.}
    \label{fig15}
\end{figure}

\begin{figure}[!ht]
    \centering
    % \subfigure[]{\includegraphics[width=0.44\textwidth]{ ncrsX7t.jpg}}
    % \subfigure[]{\includegraphics[width=0.44\textwidth]{ ncsX7t.jpg}} 
    \subfigure[]{\includegraphics[width=0.44\textwidth]{ 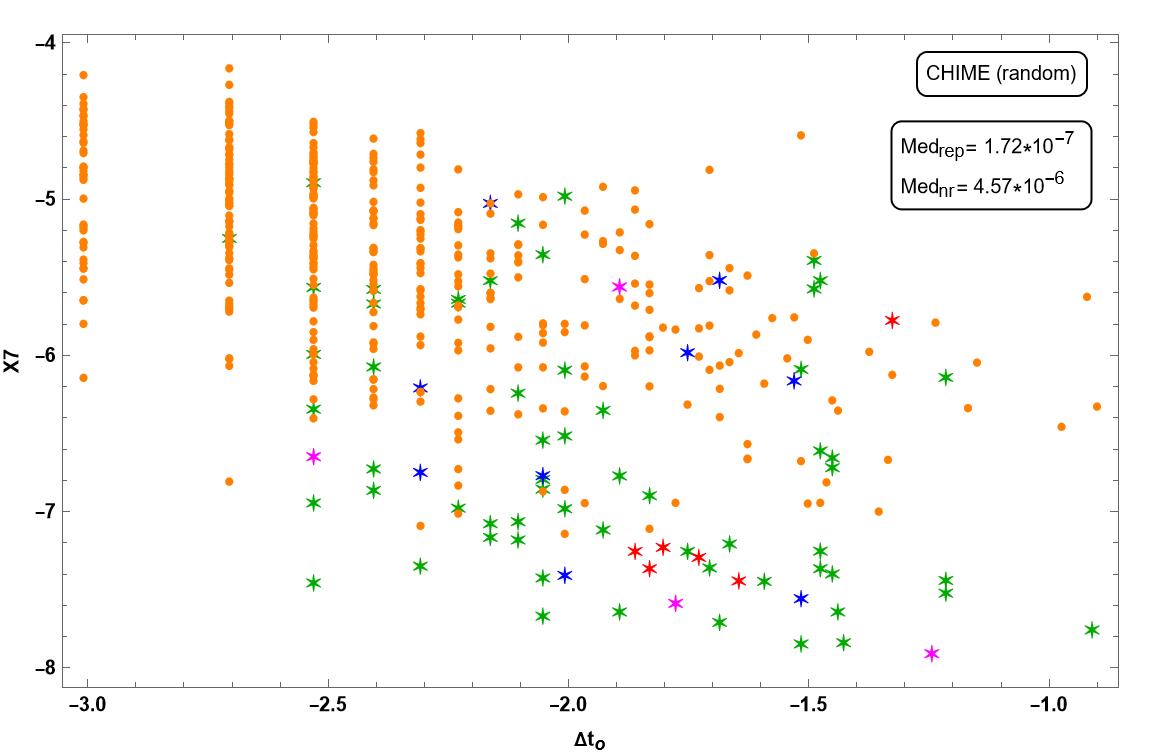}}\\
    \subfigure[]{\includegraphics[width=0.44\textwidth]{ 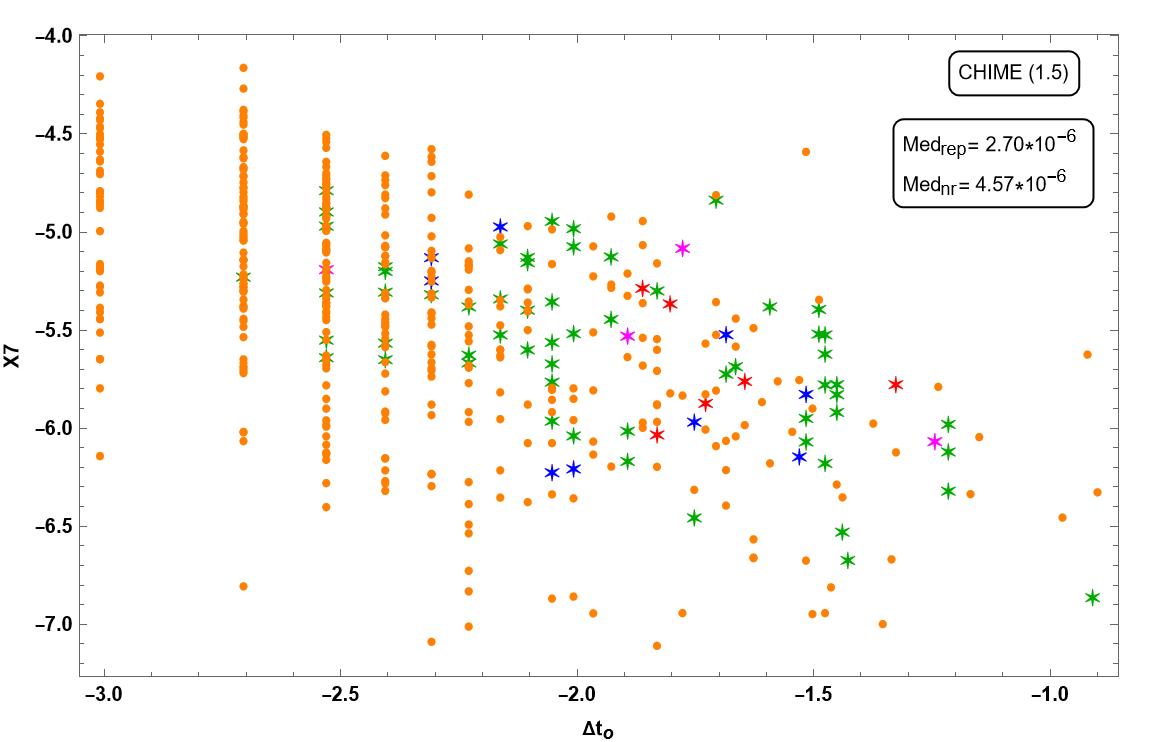}} 
    \caption{Scatter diagrams of $X7$ for CHIME FRBs against the observed FRB duration: (a) With random values of $\alpha $  assigned to the repeaters and (b) with $\alpha = 1.5$ is assigned to all the repeaters.  }
    \label{fig16}
\end{figure}

%%Use table environment for a table in one column

\end{document}